\documentstyle[aps,psfig,multicol]{revtex}  %  Swap before submission

\begin{document}
%\title{Modulational instability in periodic, quadratically nonlinear slab waveguides}
\title{Plane waves in periodic, quadratically nonlinear slab waveguides: stability and exact Fourier structure}

\author{Joel F.\ Corney and Ole Bang}
\address{Dept. of Informatics and Mathematical Modelling, Technical University of Denmark, DK-2800 Kgs. Lyngby, Denmark\\Ph: +45 45253108, Fax: +45 45931235, jfc@imm.dtu.dk}
\date{\today}
\maketitle
\begin{abstract}
We consider the propagation of broad optical beams through slab waveguides with a purely quadratic nonlinearity and containing linear and nonlinear long-period quasi-phase-matching gratings. An exact Floquet analysis on the periodic, plane-wave solution shows that the periodicity can drastically alter the growth rate of the modulational instability but that it never completely removes the instability.  The results are confirmed by direct numerical simulation, as well as through a simpler, approximate theory for the averaged fields that accurately predicts the low-frequency part of the gain spectrum.
\end{abstract}

\begin{multicols}{2}  %remove before submission

\newlength{\figwidtha}  %these are my own definitions
\newlength{\figwidthb}
\newlength{\figwidthc}
\setlength{\figwidtha}{0.45\linewidth}
\setlength{\figwidthb}{0.95\linewidth}
\setlength{\figwidthc}{0.75\linewidth}

\section{INTRODUCTION}
Optical materials with $\chi^{(2)}$ nonlinearity offer many advantages as candidates for all-optical processing applications.  For example, they offer a fast and strong nonlinear response, and through cascading $\chi^{(2)}:\chi^{(2)}$ processes \cite{SteHagTor96}, they can support stable localized solitary-wave solutions in more than one dimension \cite{HayKos93,TorMenTorSte95,BerMezRasWyl95,PelBurKiv95,MalDruetal97}, in contrast with systems that have only a cubic nonlinear response.  Experimental studies have confirmed the existence of one-\cite{SchBaeSte96} and two-dimensional\cite{TorWanetal95} spatial and spatio-temporal\cite{LiuBecWis00} quadratic solitons and have demonstrated that they act as dynamically reconfigurable guiding structures\cite{LopCouetal01}, with switching behavior controllable through phase-\cite{CouLopSimBar01} and intensity-dependent\cite{TorAssetal96} effects.  Even though special `dark-vortex' solitons have been generated in quadratic media\cite{DiTChietal00}, coherent dark solitons are generally unstable in homogeneous materials with only quadratic nonlinearity, due to a modulational instability (MI) in the background plane wave.  

Like solitons, MI occurs due to an interplay between nonlinear and dispersive or diffractive effects.  In nonlinear optics, it appears as a transverse instability that breaks up a broad optical beam\cite{TriFer95}, thereby acting as a precursor for
the formation of stable bright spatial solitons, as has been confirmed experimentally\cite{Fueetal97}.
Conversely, the stable propagation of dark solitons relies on the 
stability of the constant-intensity background and thus requires 
the absence of MI.  The mismatch\cite{LiuBecWis00} and intensity\cite{FanMalSchSte00} dependence of the instability period has been investigated experimentally, as has the intensity and frequency dependence of the gain\cite{SchFanMalSte01}. 

In all $\chi^{(2)}$ materials, there is of course an inherent cubic (Kerr) nonlinearity.  The combined nonlinearities lead to novel effects, such as the existence of bright solitons in an intrinsically defocusing cubic medium\cite{BanKivBur97}.  Also, if the defocusing cubic nonlinearity is sufficiently strong, it may eliminate MI \cite{AleBurKiv98}.  Unfortunately, the cubic nonlinearity in conventional $\chi^{(2)}$ materials is usually focusing and thus of the wrong type for these effects to occur.

However a cubic nonlinearity of a different kind may be induced through long-period variations in the linear or nonlinear susceptibilities.  Such periodicities are used for quasi-phase-matching (QPM), which compensates the material mismatch between the fundamental and second-harmonic (SH) modes.  The lowest-order effect of QPM is to induce an effectively
phase-matched quadratic nonlinearity in the equations for the averaged fields.  To first order, the non-phase-matched oscillatory components of the fields induce cubic
nonlinearities\cite{ClaBanKiv97,BanBalChrTor99,CorBan01}, in the same manner as the non-phase-matched interactions do in the cascading limit\cite{DiTBraetal01}.  For certain materials and etching techniques, altering
the nonlinear susceptibility also induces a change in the linear susceptibility.  The action of such a simultaneous linear grating
can be to reduce the effective quadratic nonlinearity\cite{FejMagJunBye92,SuhNis90,JasArvLau86,TanBey73,Pet96}, making the effective nonlinearity more cubic, as in the cascading limit.

In this paper, we investigate the effects of such periodicities on the MI properties of beams in $\chi^{(2)}$ materials, considerably extending the initial results presented in Ref.\ \onlinecite{CorBan01a}.  We provide a comprehensive and comparative treatment of the different approaches (exact and approximate) and use a variety of particular cases to illustrate general properties.  Section \ref{system} introduces the system studied in the rest of the paper.  In analyzing this system, we first (Sec.\ \ref{average_solns}) seek modulationally stable dark solitons in an approximate {\em averaged} system.  Previous work\cite{BanBalChrTor99,CorBan01} has confirmed the accuracy of the averaged
theory in describing the average properties of solitons and CW switching\cite{KobFedBanKiv98,BanGraCor01}, showing
that this intrinsically quadratically nonlinear system behaves as
though it were governed by cubic as well as quadratic nonlinearities.  The theory also describes space-varying behavior, but to a lower accuracy.  We discover that the cubic terms that result from particular gratings can be large enough and of the correct sign to prevent MI.

The second part of this work compares the stability predictions of
the averaged theory with those of exact calculations. The exact
methods involve finding the exact periodic plane-wave solutions
in the full (nonaveraged) system, with the higher harmonics
included to all orders (Sec. \ref{floquet_solns}).  We apply a Floquet analysis to these
periodic solutions to determine the growth rate of linear
perturbations (Sec.\ \ref{floquet_mi}), which can now be directly probed in experiments\cite{SchFanMalSte01}.  Additionally, we directly simulate the periodic-field equations (Sec.\ \ref{propagative}), using for initial
conditions the averaged solutions seeded with noise.

According to the exact calculations, the periodicities can drastically alter the MI gain spectrum, but they do not entirely remove the inherent instability.  In the regime of efficient QPM, we find that the MI gain spectrum contains two distinct and well-separated features with fundamentally different physical origins (see Sec.\ \ref{appI}).
The low-frequency part of the gain is accurately predicted by the averaged theory and disappears for certain grating modulations.  The high-frequency part of the spectrum is related to the inherent gain of the non-phase-matched material (i.e., with no gratings) and appears to be unavoidable.  However, because they are consistently small, the high-frequency peaks can be ignored under a less stringent definition of {\em experimental} stability (see Sec.\ \ref{exp}).  This gives the possibility of stable plane waves as predicted by the averaged theory.

\section{SYSTEM}
\label{system}

\begin{figure}
 \centerline{\hbox{  \psfig{figure=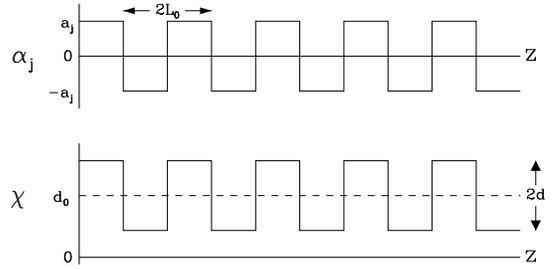,width=\figwidthb}}}
  \caption{Periodic linear and quadratically nonlinear gratings, $\alpha_j(Z)$ and $\chi(Z)$, with period $2L_0$=$2\pi/|k_g|$.} 
  \label{modulation}
\end{figure}

We consider a CW beam (carrier frequency $\omega$) interacting with its second harmonic (SH), both propagating in a lossless 1D slab waveguide under conditions for type I second-harmonic generation (SHG).  Both the linear and quadratic nonlinear susceptibilities are periodic along 
the $Z$-direction of propagation but are constant along the transverse $X$-direction.
We assume a weak modulation of the refractive index ($\Delta n_j(Z)/
\bar{n}_j$$\ll$1, where $n_j(Z)=\bar{n}_j+\Delta n_j(Z)$ and $j$ refers 
to the frequency $j\omega$).  We consider gratings for forward QPM only.
The grating period is then much longer than the optical period, in which 
case Bragg reflections can be neglected.  The evolution of the slowly 
varying beam envelopes is then described by
\cite{ArmBloDucPer62}
\begin{eqnarray}
  i\partial_Z E_1 + \frac{1}{2} \partial_X^2 E_1 + \alpha_1 E_1 + 
  \chi E_1^*E_2 \exp{(i\beta Z)} & = & 0 ,\nonumber\\
  i\partial_Z E_2 + \frac{1}{4} \partial_X^2 E_2 + 2\alpha_2 E_2 + 
  \chi E_1^2 \exp{(-i\beta Z)} & = & 0, 
  \label{field_eqns}
\end{eqnarray}
where $E_1$=$E_1(X,Z)$ and $E_2$=$E_2(X,Z)$ are the envelope functions of the 
fundamental and SH, respectively.
The coordinates $X$ and $Z$ are in units of the input beam width $X_0$ and 
the diffraction length $L_d$=$k_1 X_0^2$, respectively. 
The parameter $\beta$=$\Delta k L_d$ is proportional to the 
mismatch $\Delta k$=$k_2-2k_1$, where $k_j$=$j\omega\bar{n}_j/c$ is the 
average wave number. Thus $\beta$ is positive for normal dispersion and 
negative for anomalous dispersion.
The normalized refractive-index grating is $\alpha_j(Z)$=$L_d\omega\Delta 
n_j(Z)/c$, and the normalized nonlinear grating is $\chi(Z)$=$L_d\omega 
d_{\rm eff}(Z)/(\bar{n}_1 c)$, where $d_{\rm eff}$=$\chi^{(2)}/2$ is given 
in MKS units.  
The model (\ref{field_eqns}) describes both temporal and spatial solitons
\cite{MenSchTor94,Ban97}.  Throughout this paper, we focus on first-order QPM 
using conventional square gratings with a $50\%$ duty cycle, as shown in 
Fig.~\ref{modulation}.   

To consider plane wave solutions with longitudinal wave number offset $\Lambda$ 
and transverse wave number $\Omega$, we transform to the new variables 
$w(x,z)=E_1(X,Z)\exp{(i\Theta)}$, $v(x,z)=E_2(X,Z)\exp
(2i\Theta + i\tilde\beta Z)$, $x=\sqrt{\eta}(X+\Omega Z)$,
and $z=\eta Z$, where $\Theta(z) = \Omega X-\Lambda Z - \int \alpha_1(z) dz$ and where $\eta=|\Lambda +\frac{1}{2}\Omega^2|$ for 
$\Lambda\neq -\frac{1}{2} \Omega^2$ and $\eta$=1 otherwise.  We have also introduced a residual effective mismatch $\tilde\beta=\beta-k_g$, where $k_g$ is the spatial frequency of the gratings.  
This gives the rescaled equations
\begin{eqnarray}
  i\partial_zw + \frac{1}{2}\partial_x^2 w - rw 
        + \chi'w^*v \exp{(i\kappa z)} &=& 0 \nonumber \\
  i\partial_zv + \frac{1}{4}\partial_x^2v - \sigma v + \alpha'v
        + \chi' w^2 \exp{(-i\kappa z)} &=& 0 ,
  \label{planewave2}
\end{eqnarray}
where $r = {\rm sgn}(\Lambda +\frac{1}{2}\Omega^2)$, $\kappa = k_g/\eta$, and $\sigma = 2r -\tilde\beta'$, for $\tilde\beta' = \tilde\beta/\eta$.
Note that $r$=0 when $\Lambda$=$-\frac{1}{2}\Omega^2$.  This scaling by $\eta$ may be absorbed into the original normalization [Eq.\ (\ref{field_eqns})] through a redefinition of $X_0$ (and hence $L_d$) to be something other than the true beam width (diffraction length).  We choose to retain $\eta$ explicitly, however, to show the connection with the standard normalization. 

To find solutions in this periodic system, we Fourier-expand the rescaled gratings, $\alpha'(z)$=$2(\alpha_2(Z) - \alpha_1(Z))/\eta$ and $\chi'(z)=
\chi(Z)/\eta$:
\noindent\begin{equation}
  \label{FourGrating}
  \alpha' = a' \sum_n g_n \exp{(in\kappa z)} , \;
  \chi'     = d_0' + d' \sum_n g_n \exp{(in\kappa z)},
\end{equation}
where $a'=2(a_2 - a_1)/\eta$ and where $g_n=2s/(i\pi n)$ for $n$ odd and $g_n=0$ for $n$ even.
We include $s={\rm sgn}(\kappa)$ in the Fourier expansions to ensure that the 
expansions give the same physical grating form for both signs of $\kappa$.  
Plane waves in system (\ref{field_eqns}) and (\ref{planewave2}) have an intensity that is constant in the 
transverse direction but periodic in $Z$, driven by the 
periodic variation of the susceptibilities. Thus they have the same 
spatial frequency $\kappa$ as the gratings, and we expand them in Fourier series also:
\noindent\begin{equation}
  \label{FourField}
  w = \sum_n w_n(x,z) \exp{(in\kappa z)} , \;
  v = \sum_n v_n(x,z) \exp{(in\kappa z)}.
\end{equation}
An accurate treatment of the problem requires that all Fourier components (up to some cutoff) be included.  This we do in Sec.\ \ref{floquet_solns}.  However we first present an approximate theory for the DC components $(\bar w,\bar v)=(w_0,v_0)$, which suffices for many calculations.

\section{AVERAGED-FIELD THEORY}
\label{average_solns}
\subsection{Plane-wave solutions}

Three physical length scales are in play: the diffraction length $L_d$, 
the coherence length $L_c$, and the grating domain length $L_0$. 
In our rescaled units $L_d=\eta$, $L_c=\eta\pi/|\beta| \equiv \pi/|\beta'|$, and $L_0=\pi/|\kappa|$.  We assume a typical QPM grating with a domain length that is much shorter 
than the diffraction length and use perturbation theory\cite{ClaBanKiv97} with 
the small parameter $\epsilon = L_0/L_d \ll 1$.  We insert the Fourier
expansions (\ref{FourGrating}) and (\ref{FourField}) into the dynamical 
equations and assume the harmonics $w_{n\ne0}$ and $v_{n\ne0}$ to be of
order $\epsilon$. Furthermore, efficient phase matching is assumed, with the domain length being 
close to the coherence length, $L_0 \simeq L_c$, so the residual 
mismatch is small, $|\tilde\beta'| \ll |\kappa|$.  To first order ($\epsilon^1$), this gives the harmonics
\begin{eqnarray}
  w_{n\ne0} &=& \frac{(d'g_{n-1}+d'_0\delta_{n,1})w_0^*v_0}{n\kappa}, \nonumber\\
  v_{n\ne0} &=& \frac{a'g_nv_0 + (d'g_{n+1}+d'_0\delta_{n,-1})w_0^2}{n\kappa},
\label{components}
\end{eqnarray}
where $\delta_{n,m}$ is Kronecker's delta and where we have assumed that the coefficients $w_n(x,z)$ and $v_n(x,z)$ vary 
slowly in $z$ compared to $\exp(i\kappa z)$.
Using these solutions, we obtain to first order the averaged 
equations for the DC components $(\bar w,\bar v)=(w_0,v_0)$:
\begin{eqnarray}
   &&i\partial_z \bar w + \frac{1}{2}\partial_x^2\bar w - r\bar w 
     + \rho \bar w^*\bar v + \gamma(|\bar v|^2-|\bar w|^2)\bar w = 0, 
     \nonumber\\
   &&i\partial_z \bar v + \frac{1}{4}\partial_x^2 \bar v - \sigma\bar v 
     + \rho^* \bar w^2 + 2\gamma|\bar w|^2\bar v = 0,
  \label{average_eqns}
\end{eqnarray}
These equations also describe $m$th order QPM (where $\tilde\beta = \beta-m k_g$
is ideally zero) and any other type of long-period grating.  Incorporating time or the spatial $y$ coordinate is also straightforward.  The effective quadratic ($\rho$) and induced cubic ($\gamma$) nonlinearities are simply given by sums over the Fourier coefficients\cite{ClaBanKiv97}, and for the square grating (\ref{FourGrating}) are $\rho= i2s \left\{d' - d_0'a'/\kappa\right\}/\pi$ and $\gamma= \left[d_0'^2+d'^2(1-8/\pi^2)\right]/\kappa$.

From Eqs.~(\ref{average_eqns}) follows the important result that
cubic nonlinearities are induced by the nonlinear QPM gratings.
This cubic nonlinearity has the form of asymmetric self- and cross-phase modulations (SPM and XPM) and is a result of non-phase-matched coupling between 
the wave at the main spatial frequency $\kappa$ and its higher harmonics. 
It is thus of a fundamentally different nature than the material Kerr nonlinearity, a fact which is reflected in the the SPM being absent for the SH.

The induced cubic nonlinearity 
depends only on the nonlinear 
grating and may lead to either a focusing or a defocusing effect, 
depending on the relative intensity of the fields and the sign of the 
phase mismatch $\beta'$, since ${\rm sgn}(\kappa)= {\rm sgn}(\beta')$.  
The effective $\chi^{(2)}$ nonlinearity has a contribution from the linear grating, which can either reinforce or undermine the contribution from the nonlinear grating, depending on the physical situation 
\cite{TanBey73,JasArvLau86}. 

Any plane wave  will correspond to a stationary solution $(\bar w_s, \bar v_s)$ of Eq.~(\ref{average_eqns}), obtained by setting the derivatives to zero.  By using a gauge transformation, we make $\bar w_s$ real and positive.  The solutions are then $\bar w_s = \sqrt{y}/|\rho|$, $\bar v_s = y/(\sigma\rho - 2\tilde\gamma y\rho)$, where $y$ is a positive, real number that satisfies 
\begin{equation}
0 = 4\tilde\gamma^3y^3 + \tilde\gamma\left(1+4\tilde\gamma(r-\sigma)\right) y^2 + \sigma\left(\tilde\gamma\sigma -4r\tilde\gamma -1\right)y +r\sigma^2. \label{x_solutions}
\end{equation}
For a given $r$, $\tilde \gamma$ and $\sigma$, there can thus be anywhere up to three solutions.  The relative strength of the cubic nonlinearity is given by the parameter $\tilde\gamma = \gamma/|\rho|^2$, which depends only on the material grating parameters $\kappa$, $a'/\beta'$, and $d_0'/d'$.  Note that the residual-mismatch term $\sigma$ equals $2r$ for perfect phase matching. 

\begin{figure}
\centerline{ \psfig{figure=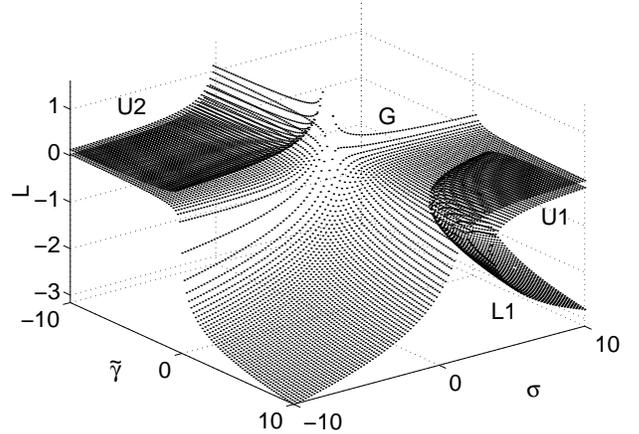,width=\figwidthb}}
\caption{Ratio of intensities $L = \log(\bar v_s^2 / \bar w_s^2)$ for plane-wave solutions, as a function of $\tilde\gamma$ and $\sigma$, for $r = -1$} 
\label{plane_solns}
\end{figure}
 
The solutions in parameter space ($\tilde \gamma, \sigma$) are illustrated by the intensity ratio $R = \bar v_s^2/\bar w_s^2$ in Fig.\ \ref{plane_solns} for $r = -1$. The figure shows that there are four extended branches of solution, which exist in three quadrants, the fourth quadrant being devoid of any solution. We label the pervasive branch (G) and the upper branch in the 
%($\tilde\gamma < 0$,$\tilde\gamma < 0$) 
negative quadrant (U2).  The boundaries of (G) and (U2) lie along the $\tilde\gamma$ and $\sigma$ axes and are where the ratio $R$ tends to infinity.  The positive quadrant contains additional paired upper (U1) and lower (L1) branches, which exist only away from the axes, their boundary given by:
\begin{eqnarray}
\sigma(\tilde\gamma)  &=& \left[5+16\tilde\gamma + 3(\sqrt[3]f_- + \sqrt[3]f_+)\right]/(8\tilde\gamma), \label{boundary}
\end{eqnarray} 
where $f_\pm = -1 + 80\tilde\gamma + 128\tilde\gamma^2 \pm 16\sqrt{\tilde\gamma(-1+4\tilde\gamma)^3}$.  Note that for $\tilde \gamma <1/4$, we must take the roots of $f_\pm$ that give the largest real $\sigma$.  Equation (\ref{boundary}) tells us that (U1) and (L1) exist only for $\tilde\gamma > 0$ and $\sigma > 2$.

Note that the $r = -1$ solutions can be transformed to the $r = 1$ solutions by the replacement $\bar v_s \rightarrow -\bar v_s$, $\sigma \rightarrow -\sigma$ and $\tilde\gamma \rightarrow -\tilde\gamma$. However, the MI properties differ.  In addition to the solutions in Fig.\ \ref{plane_solns}, there is also a trivial solution $(\bar w_s,\bar v_s)=(0,0)$ and a degenerate solution $(\bar w_s,\bar v_s)=(0,{\rm arbitrary})$ that exists for $\sigma = 0$.  However, the degenerate solution does not correspond to any dark soliton solution, even though it is sometimes stable, as there is no nonlinearity in the equation for $\bar v_s$ when $\bar w_s=0$.

\begin{figure}
\centerline{\hbox{ \psfig{figure=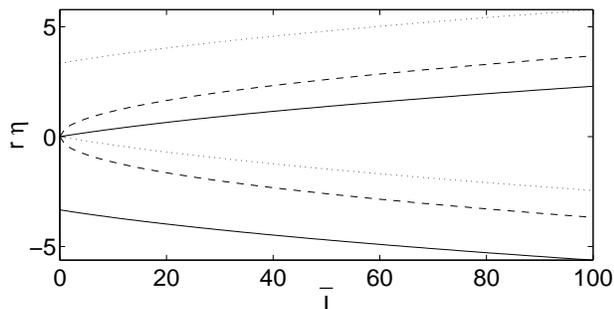,width=\figwidthb}}}
\caption{Parameter $r\eta$ versus averaged intensity $\bar I$ for $\tilde\beta = -10$ (solid), $\tilde\beta = 0$ (dashed), $\tilde\beta = 10$ (dotted), with  $d_0 = 0$, $d=1$ and $k_g = 100$.} 
\label{nl_disp}
\end{figure}

To relate the solutions (\ref{x_solutions}) to a specific physical situation requires that the plane-wave variables $r$ and $\eta$, and hence $\Omega$ and $\Lambda$, be known, since $r$ specifies the solutions and $\eta$ is scaled into the normalized units and grating parameters. Now $\Omega$ is determined by the angle of the plane wave to the $Z$-axis and is zero at normal incidence.  In contrast, $\Lambda$ is not directly accessible in experiments; it depends on the intensities of the fundamental and SH and must be determined through a nonlinear dispersion relation.  In the limit $\tilde\gamma \ll 1$, the cubic terms can be ignored, which gives $r = (\tilde \beta' \pm \sqrt{\tilde\beta'^2 + 12 \bar I |\rho|^2})/6$, where $\bar I = |\bar w|^2 + |\bar v|^2$ is the averaged intensity. In unscaled units $\eta r = (\tilde \beta \pm \sqrt{\beta^2 + 12 \bar I |\rho_{\rm UN}|^2})/6$, where $\rho_{\rm UN} = \eta \rho$ has now no dependence on $\eta$.  Which root of the radical is taken depends on the relative size of the two field intensities.  When cubic terms need to be included, a simple analytic expression for $r$ or $r\eta$ is not possible and the calculation must be done numerically.  As an example, Fig.\ \ref{nl_disp} illustrates the dependence of $r\eta$ on $\bar I$ and $\tilde \beta$ for a symmetric nonlinear grating.  The two branches of solution are distinguished by different ratios of the fundamental to SH.  Where a choice of $\eta$ is required in Secs. \ref{floquet_solns}-\ref{propagative}, we simply take $\eta = 1$, which can be satisfied by an appropriate choice of intensity $\bar I$ and input width $X_0$. 
 
\subsection{Modulational instability}
\label{average_mi}
Following standard MI analysis\cite{AleBurKiv98,HeDruMal96,HeArretal99}, we assume that the plane-wave solutions $\bar w_s$ and $\bar v_s$ of Eq.~(\ref{average_eqns}) are perturbed by small modulations:
\begin{eqnarray}
\bar w (z) &=& \bar w_s + \bar\delta_1(z)\exp{(- i \nu x)} +  \bar\delta_2^*(z)\exp{(i \nu x)} \nonumber \\
\bar v (z) &=& \bar v_s + \bar\delta_3(z)\exp{(- i \nu x)} + \bar\delta_4^*(z)\exp{(i \nu x)}, 
\end{eqnarray} 
where $*$ denotes the complex conjugate.  The linear growth is governed by $\partial_z{\bf \bar \Delta}(z) = \bar M{\bf \bar \Delta}(z)$, where  
%${\bf \bar \Delta}  = \left[\bar\delta_1 ,\bar\delta_2  ,\bar\delta_3  ,\bar\delta_4\right]^T$ and the constant matrix $\bar M$ is
\begin{eqnarray}
{\bf \bar \Delta} = \left(\begin{array}{c}
\bar\delta_1\\
\bar\delta_2 \\
\bar\delta_3 \\
\bar\delta_4 
\end{array}\right),\;
\bar  M = i\left(\begin{array}{cccc}
a & b & c & d \\ -b & -a & -d & -c \\ 2c & 2d & g & 0 \\ -2d & -2c & 0 & -g \end{array}\right),
\label{stab_mat}
\end{eqnarray}
with $ a = -\nu^2/2 - r + \gamma(|v_s|^2 - 2w_s^2)$, $b = \rho v_s -  \gamma w_s^2$, $c = \rho w_s +  \gamma v_s w_s$, $d = \gamma v_s w_s$ and $g = -\nu^2/4 - \sigma + 2\gamma w_s^2$.  For MI to be absent, all the eigenvalues $\lambda_i$ of $\bar M$ must be imaginary.  In practice, we calculate the gain at a finite number of values of $\nu$, out to a maximum of $\nu = 50$.  

\begin{figure}
\centerline{\psfig{figure=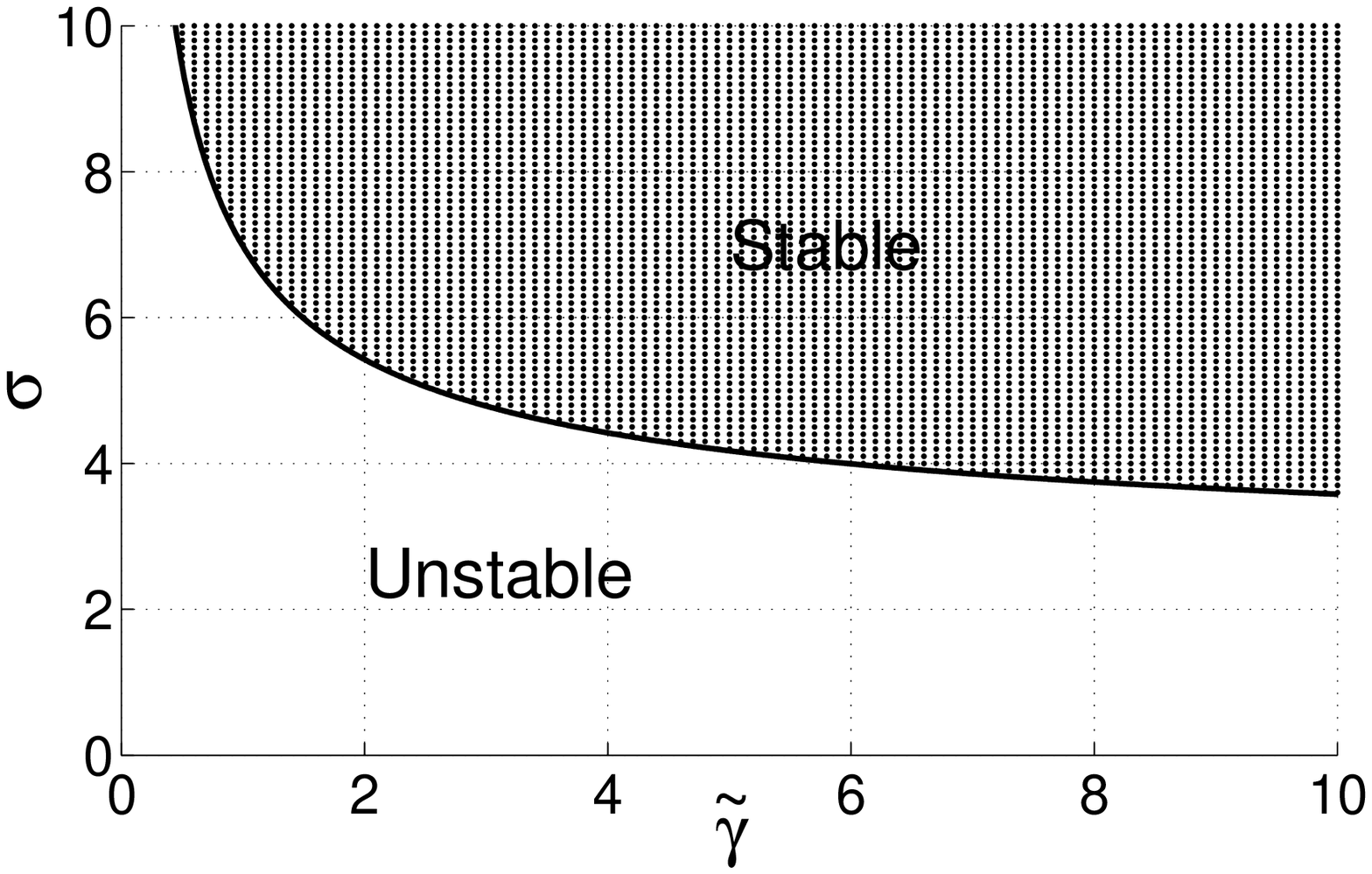,width=\figwidthb}}
\caption{(a) Region of stability for the $r=-1$ solutions.} 
\label{stability}
\end{figure}

For $r = 1$ and $r = 0$, stability analysis shows that {\em all} nondegenerate and nontrivial solutions are {\em unstable}.  For the $r=-1$ solutions, however, the (L1) branch of solutions is stable.  Figure \ref{stability} shows the region of stability, which differs significantly from that found in %Alexander, Buryak and Kivshar
Ref.\ \onlinecite{AleBurKiv98} because of the asymmetric nature of the induced cubic nonlinearity in Eq.\ (\ref{average_eqns}).  The continuous line in Fig.\ \ref{stability} is the boundary of (L1).  Now on this branch, $\tilde\gamma$ and $\sigma$ cannot both be small, suggesting that there are two possible ways to achieve stability.  One is to have a relatively large cubic nonlinearity $\tilde\gamma$, and the other a large mismatch $\sigma$.  Increasing the size of $\tilde\gamma$ basically entails decreasing $\rho$, the size of the effective quadratic nonlinearity, if the perturbation theory is to remain valid. However, in most applications a large effective $\chi^{(2)}$ is desirable.  The other possible route to stability is to increase $\sigma$, i.e. to increase $-\tilde\beta' = -\tilde\beta/\eta$.  However, this large residual mismatch may also invalidate the perturbation theory. ($\eta$ cannot be reduced, because this would decrease $\tilde\gamma$, thereby moving the solution further from the stable region.)  Thus whether this region of stability truly exists must be checked by more exact methods. 

\begin{figure}
\centerline{\hbox{ \psfig{figure=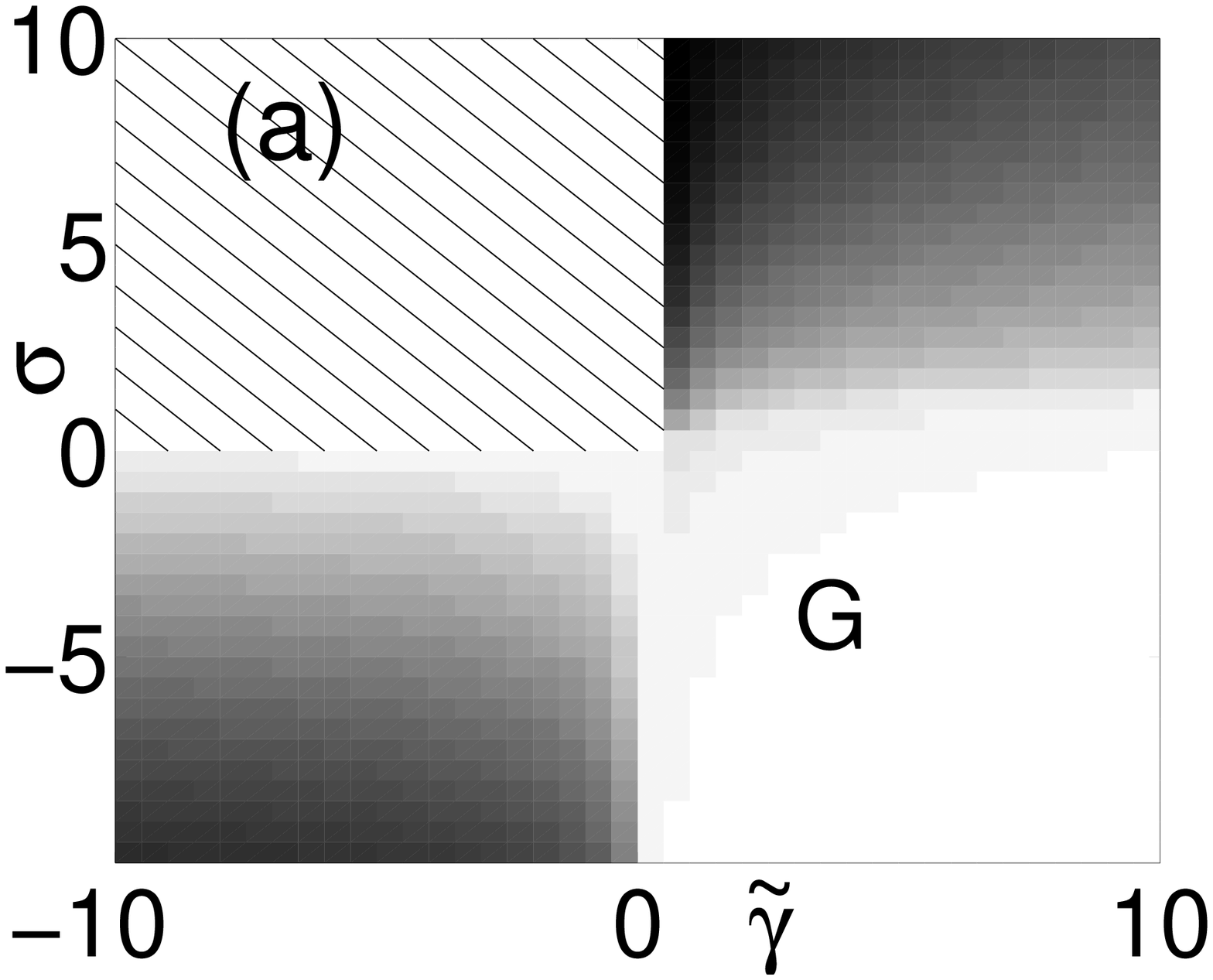,width=\figwidtha} \psfig{figure=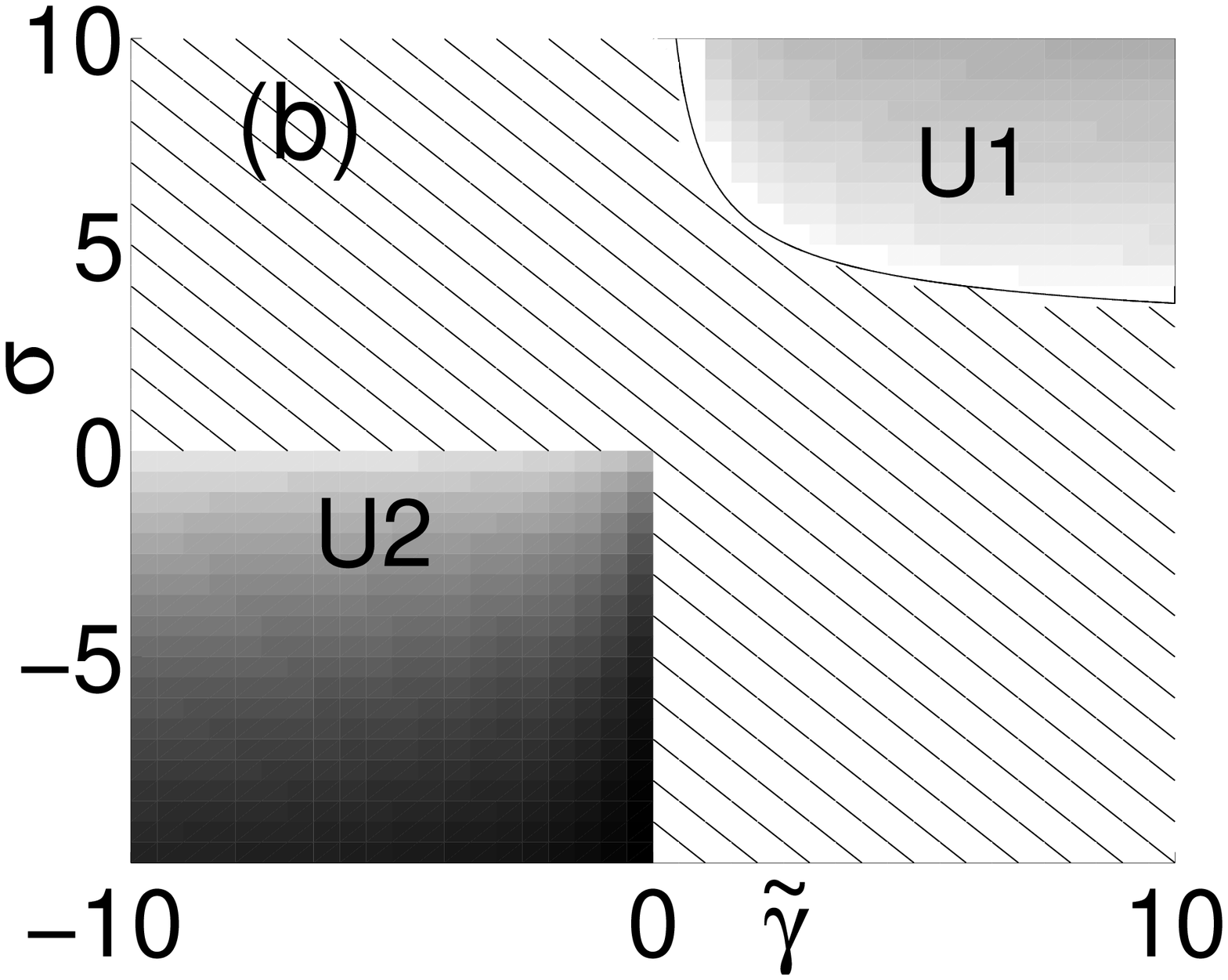,width=\figwidtha}}}
\caption{Maximum gain of the unstable branches (a) G and (b) U1 and U2 of the $r=-1$ solutions (Black is largest gain).  The hatching indicates the absence of a solution.} 
\label{unstable}
\end{figure}

Figure \ref{unstable} plots the maximum gain of the unstable branches (G,U1,U2).  It shows several regions where the gain is small, namely close to the existence boundaries of the solutions (G,U1,U2) and also for (G) when $\tilde\gamma \gg 1$, $\sigma \ll -1$.  Most of these `almost stable' regions correspond to either $\bar w_s$ or $\bar v_s$ becoming small.  For example, near  $\sigma = 0$, the solutions (G) and (U2) approach the degenerate solution ($\bar w_s = 0$) for $\tilde\gamma < 1/4$; and when $\tilde\gamma \gg 1$, $\sigma \ll -1$, the SH in (G) approaches zero.  In these cases, the averaged equations effectively become either linear or a NLS equation.  The exception is the branch (U1): its gain becomes small in the region near where it meets the stable (L1) branch, as expected.  Now some of these solutions have such a low gain that for experimental purposes they will appear to be stable.  Figure \ref{unstable} tells us that such {\em experimentally stable} solutions lie in diverse regions in parameter space and thus may be more accessible than the mathematically stable solutions.

\begin{figure}
\centerline{\hbox{ \psfig{figure=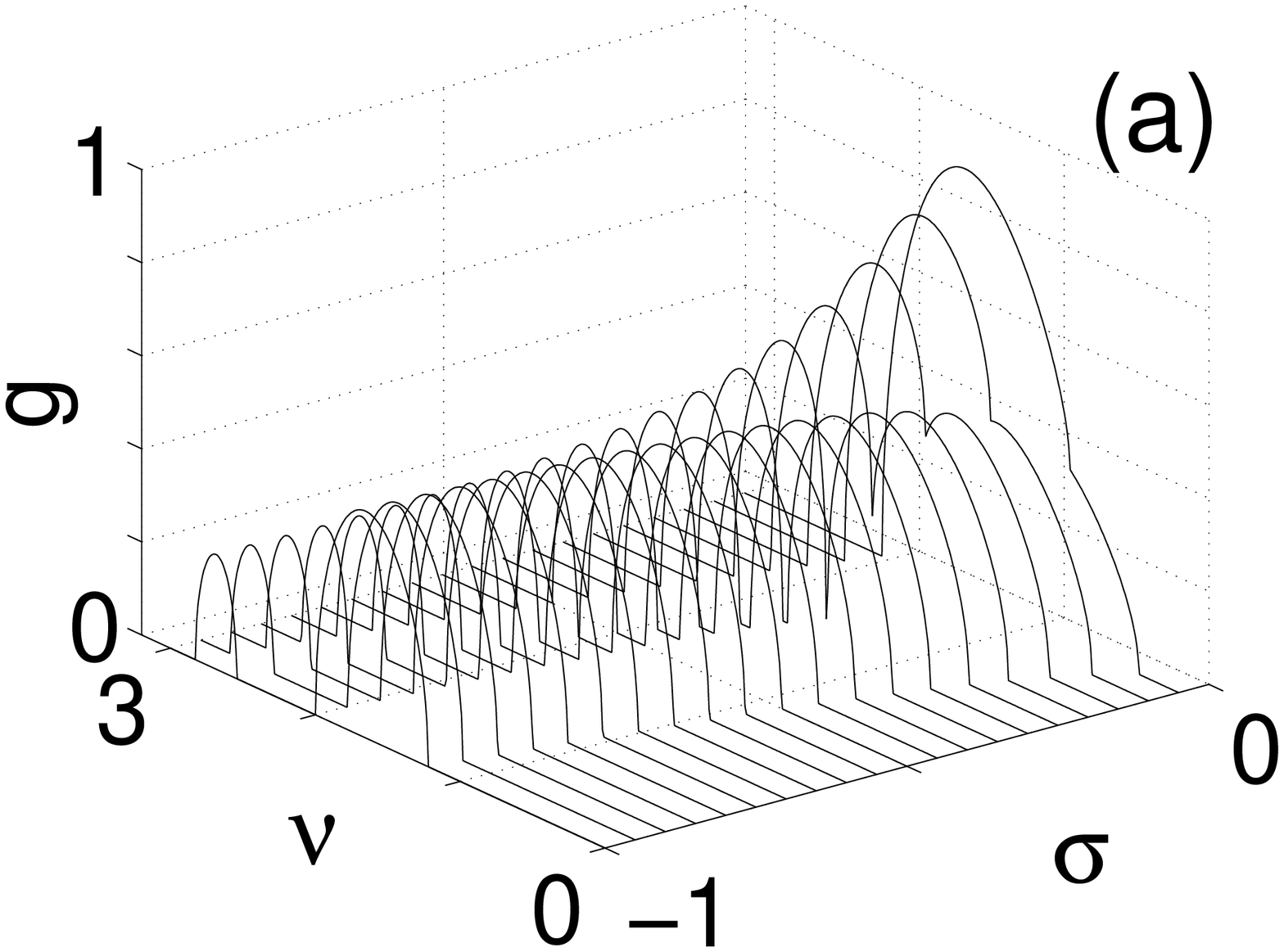,width=\figwidtha} \psfig{figure=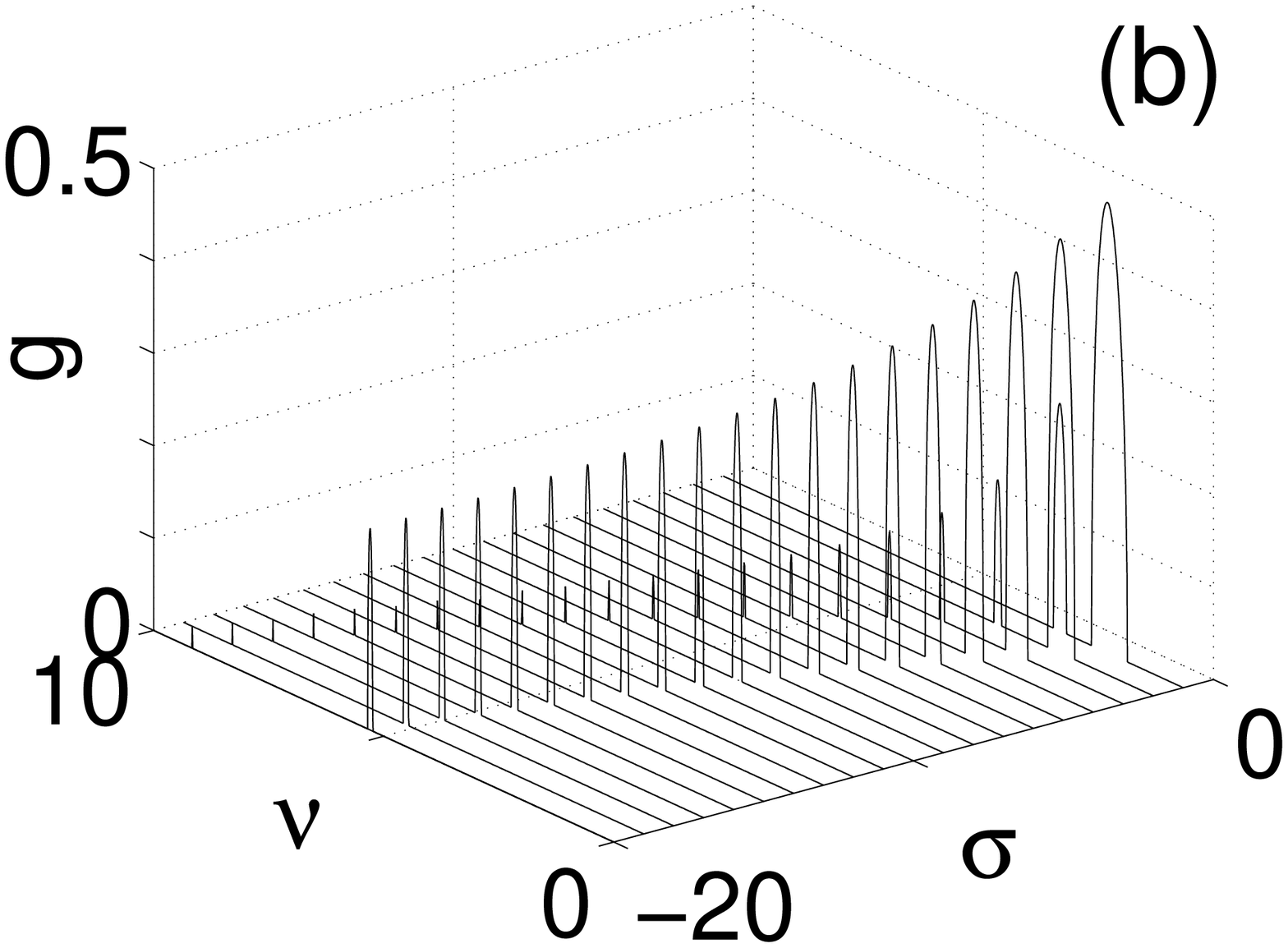,width=\figwidtha}}}
\caption{Gain profiles of the $r=-1$ solutions for $\tilde\gamma = 0$.} 
\label{gain_curve}
\end{figure}

We show in Fig.\ \ref{gain_curve} the gain profiles for $\tilde\gamma = 0$ of the (G) solution, which corresponds to when the induced cubic terms are ignored, i.e. when the first-order terms are not taken into account in the perturbation theory.  The same gain curves appear when there is no grating, just with the effective mismatch $\tilde\beta$ replaced by the (very much larger) inherent mismatch $\beta$ in the expression for $\sigma$, which gives a larger $|\sigma|$.  Thus to zeroth order, the effect of the periodic grating is to increase and broaden the gain, because a much smaller $|\sigma|$ is achieved than in the non-phase-matched case without a grating.  However, the more accurate first-order theory involves a further adjustment to the spectrum, with the inclusion of the cubic terms (so that $\tilde\gamma \neq 0$).  The cubic terms may decrease the maximum gain, perhaps even to zero for particular gratings.

\subsection{Cubic solutions: $\rho = 0$}
\label{cubic_section1}
For some combinations of linear and nonlinear gratings, the cubic nonlinearity may dominate the quadratic. In the extreme case, the effective quadratic nonlinearity is zero ($\rho = 0$), caused by the linear grating (multiplied by the strength of the DC nonlinearity) cancelling the contribution of the nonlinear grating, which can occur for realistic physical parameters.  The averaged-field equations reduce to two NLS equations, which support the stable dark solitons (for $r = -1$): 
\begin{eqnarray}
  \bar w &  = &  {\rm tanh}(x)/\sqrt{\gamma}, \; \bar v  = 0\: (\gamma > 0)\label{1dark}\\
  \bar w &=& {\rm tanh}(x)/\sqrt{|4\gamma|}, \; \bar v  = \pm\sqrt{5/|4\gamma|} {\rm tanh}(x)\:(\gamma < 0),\label{2dark}
\end{eqnarray}
where $\sigma = -1/2$ in the second solution (\ref{2dark}).  In the absence of the parametric term, the two averaged fields are coupled only by  phase-independent XPM terms, which is why these solutions do not possess the `natural' $\chi^{(2)}$ phase relation:  in the first solution (\ref{1dark}), $\sigma$ is arbitrary, and the second solution (\ref{2dark}) only exists for a particular, nonzero value of the residual mismatch $\tilde\beta' = -3/2$.  

The plane-wave backgrounds of the solitons (\ref{1dark}) and (\ref{2dark}) form part of a larger set of plane-wave solutions that exist in the absence of the quadratic terms: $(\bar w_s,\bar v_s) = (\sqrt{-r/\gamma},0)$ and $(\bar w_s,\bar v_s) = (\sqrt{\sigma/2\gamma},\pm\sqrt{(2r+\sigma)/2\gamma})$.  The $\bar v_s = 0$ solution is stable for $\gamma > 0$ ($r = -1$)  and unstable for  $\gamma < 0$ ($r = 1$), whereas the $\bar v_s \neq 0$ solution is always unstable.

\begin{figure}
\centerline{\hbox{ \psfig{figure=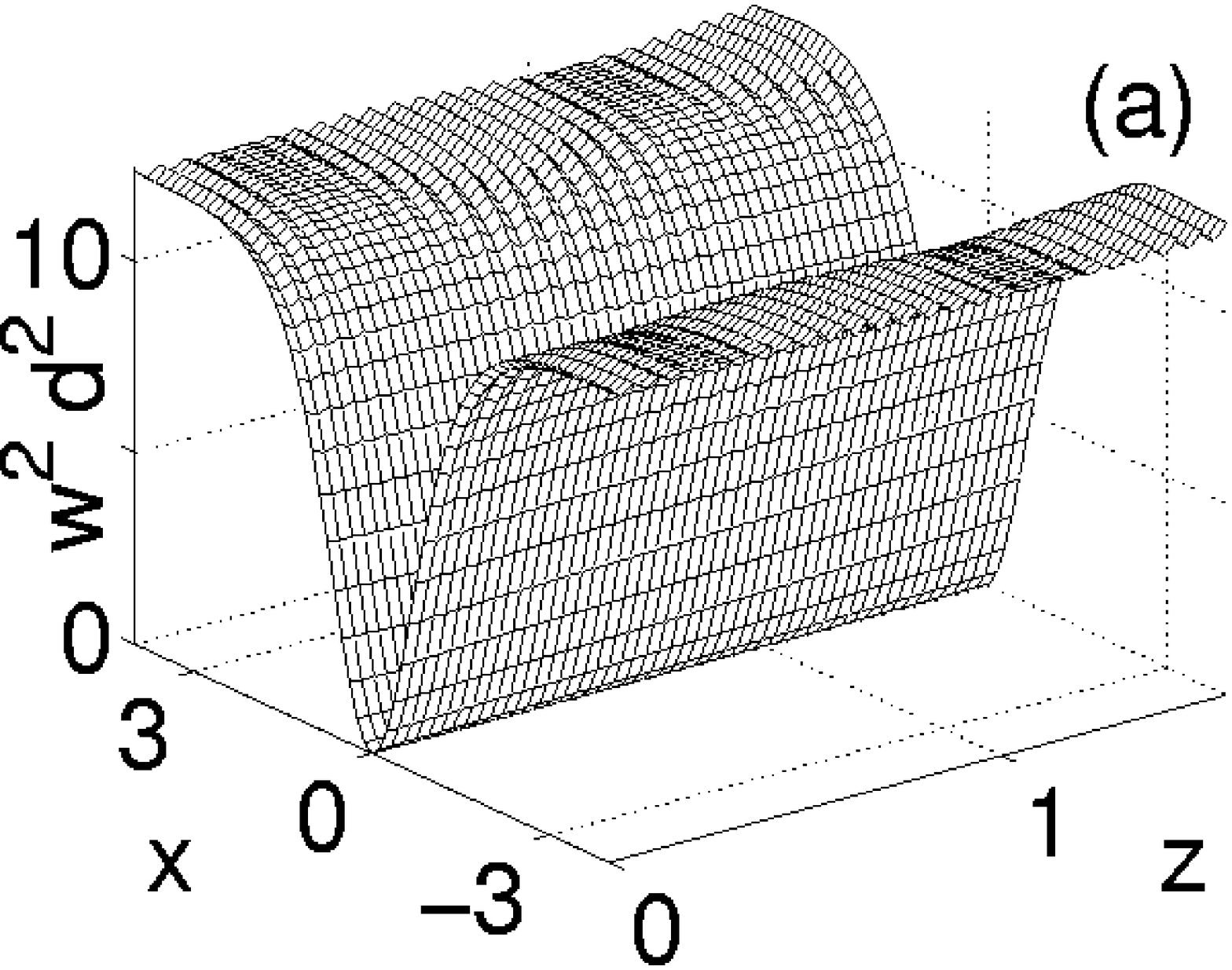,width=\figwidtha } \psfig{figure=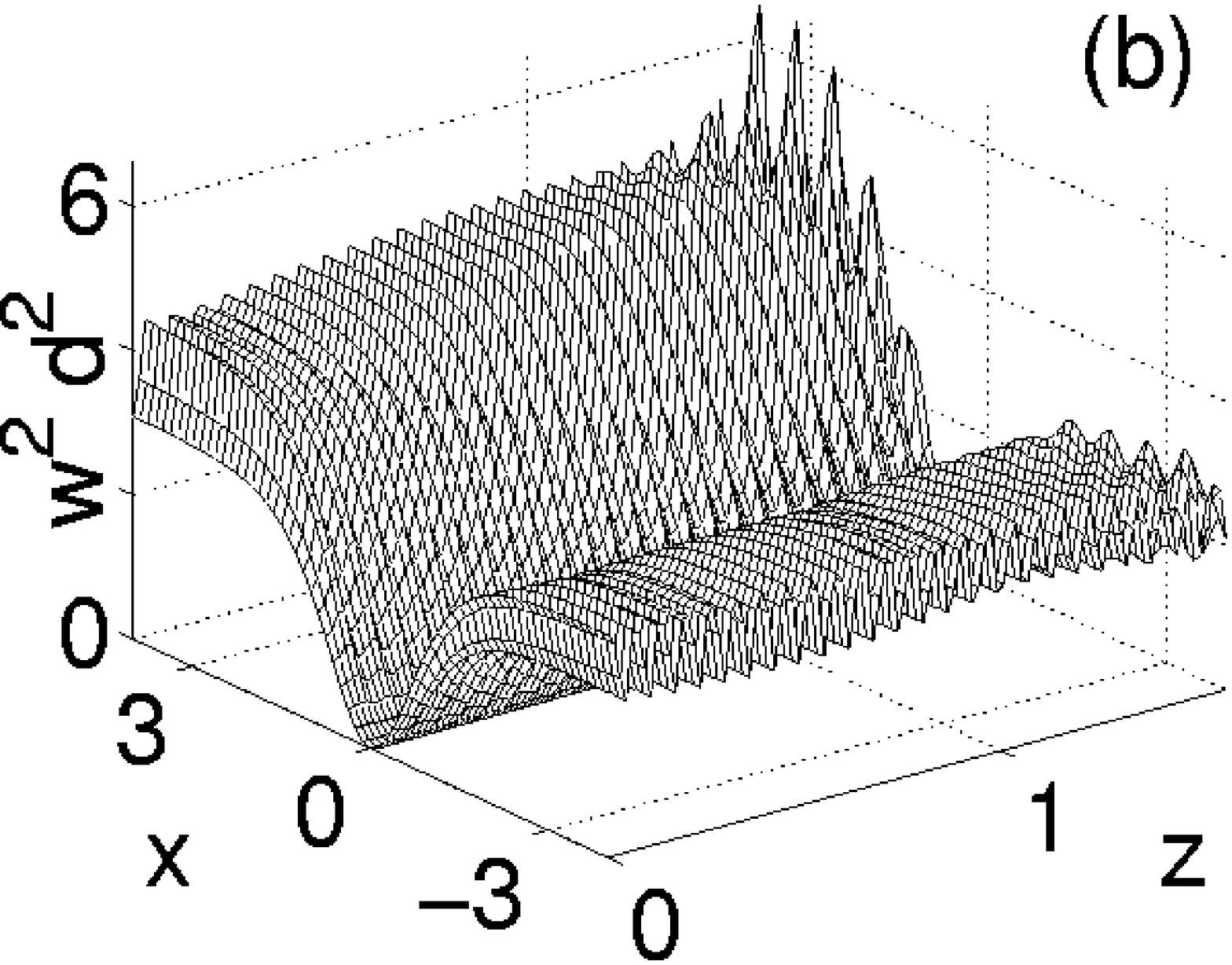,width=\figwidtha }}}
\caption{The evolution of the fundamental intensity $w$, corresponding to the solutions (a) Eq.\ (\ref{1dark}) with $\gamma = 0.08$ and (b) Eq.\ (\ref{2dark}) with $\gamma = -0.08$.  The gratings parameters are $d_0'/d' = 2.8$ and $a'/\kappa = 0.36$. r = -1.}
\label{dark1}
\end{figure}

Figure \ref{dark1}(a) shows a simulation of the original field equations (\ref{field_eqns}) with initial condition corresponding to the stable NLS soliton (\ref{1dark}).  No visible shedding of radiation occurs, and the fundamental intensity contains only rapid, small oscillations superimposed on the large average beam, which is typical for stable, bright QPM solitons\cite{CorBan01,ClaBanKiv97}.  The soliton propagates stably beyond $z=150$, which corresponds to 150 diffraction lengths or over 2000 coherence lengths.  Thus the soliton is stable according to any reasonable experimental definition.  In contrast, simulations of the two-wave dark soliton (\ref{2dark}) over short distances (Fig.\ \ref{dark1}(b)) confirm that it is indeed modulationally unstable.  We note that a small modulation (of spatial  frequency $\nu \simeq 10$) finally develops in Fig.\ \ref{dark1}(a), which indicates a weak instability not picked up by the averaged-field equations.

\section{EXACT PLANE-WAVE SOLUTIONS}
\label{floquet_solns}
The stability analysis with the averaged-field equations found regions of stability.  But these regions possibly lie at or beyond the limits of validity of the first-order perturbation theory.  Additionally, propagation simulations showed weak instabilities not predicted by the averaged theory.  Clearly, a thorough investigation requires a more exact technique that does not rely on the perturbation theory.
Therefore, we numerically find the exact plane-wave solutions and apply an exact Floquet analysis\cite{CreDauRufTor98,Fueetal97} to determine their stability. We begin with Eqs.\ (\ref{planewave2}), substituting in the Fourier expansions (\ref{FourGrating}) and (\ref{FourField}).  This gives a set of coupled equations for the plane-wave coefficients $w_n$ and $v_n$: 
\begin{eqnarray}
  -(n\kappa+r) w_n + \sum_{l,p}D_{n+l-p-1}w_l^*v_p
  &=&0, \nonumber \\
  -(n\kappa+\sigma) v_n + a'\sum_lg_{n-l}v_l + \sum_{l,p}D_{n-l-p+1}w_lw_p
  &=&0,
  \label{F}
\end{eqnarray}
where $D_n$=$d'g_n+d_0'\delta_{n,0}$.  
We may rewrite these equations more succinctly as ${\bf F}_n = 0$.
 
We find the solution to Eq.\ (\ref{F}) numerically, using a relaxation technique based on Newton's method.  An initial guess of the components $w_n$ $v_n$ (which we represent by the vector ${\bf u}^{(0)}$), calculated from the approximate first-order solutions, is iteratively corrected by the increment $\Delta {\bf u}^{(k+1)} = -J^{-1}({\bf u}^{(k)}){\bf F}({\bf u}^{(k)})$, where the Jacobean matrix is
\begin{eqnarray}
\{J\}_{l,q} = \left(\begin{array}{cccc}
\frac{\partial F_l}{\partial w_q}& \frac{\partial F_l}{\partial v_q}& \frac{\partial F_l}{\partial w_q^*}& \frac{\partial F_l}{\partial v_q^*}\\ \frac{\partial F^*_l}{\partial w_q}& \frac{\partial F^*_l}{\partial v_q}& \frac{\partial F^*_l}{\partial A^*_q}& \frac{\partial F^*_l}{\partial B^*_q}
\end{array} \right). 
\end{eqnarray}

\begin{figure}
\centerline{\vbox{\hbox{ \psfig{figure=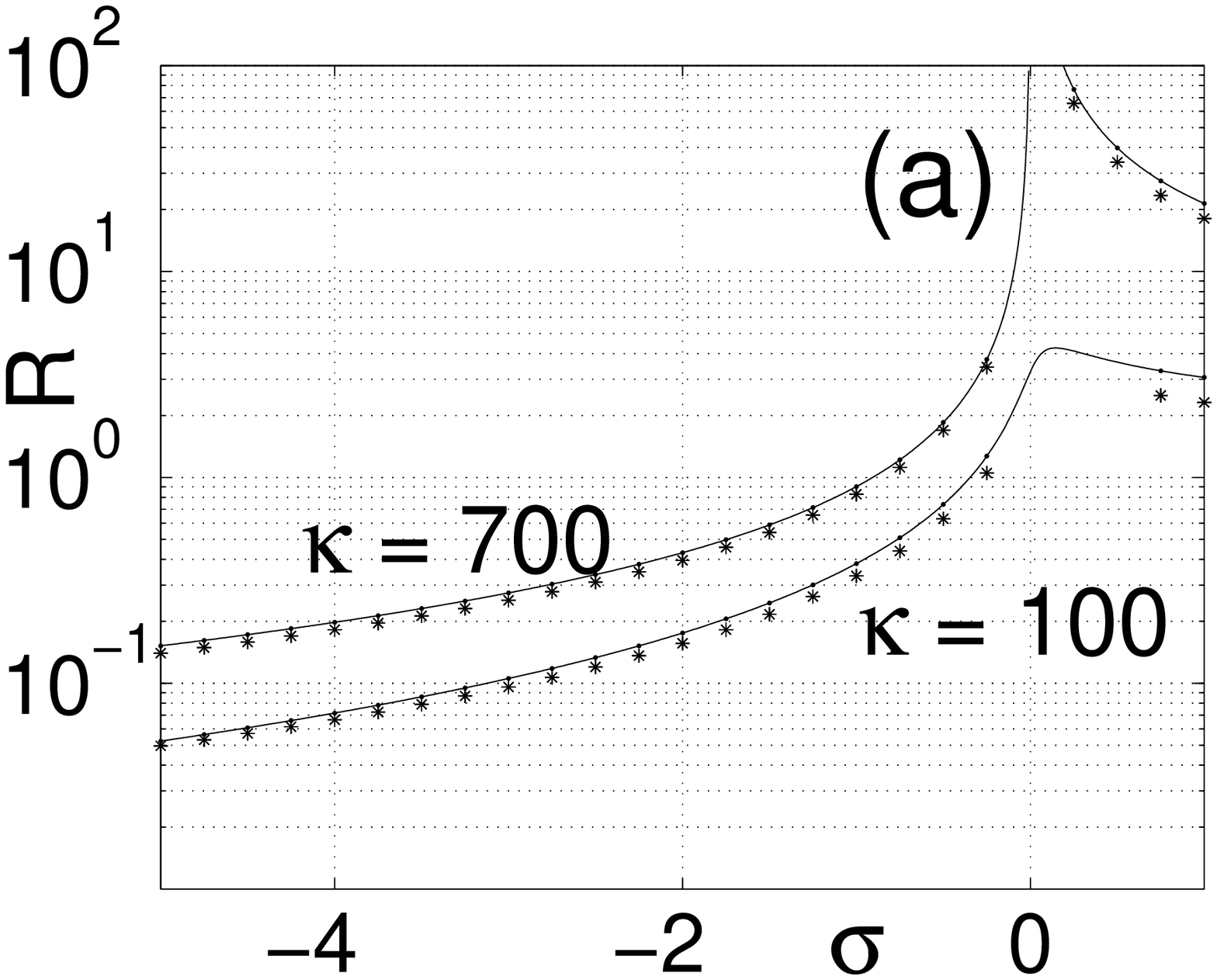,width=\figwidtha } \psfig{figure=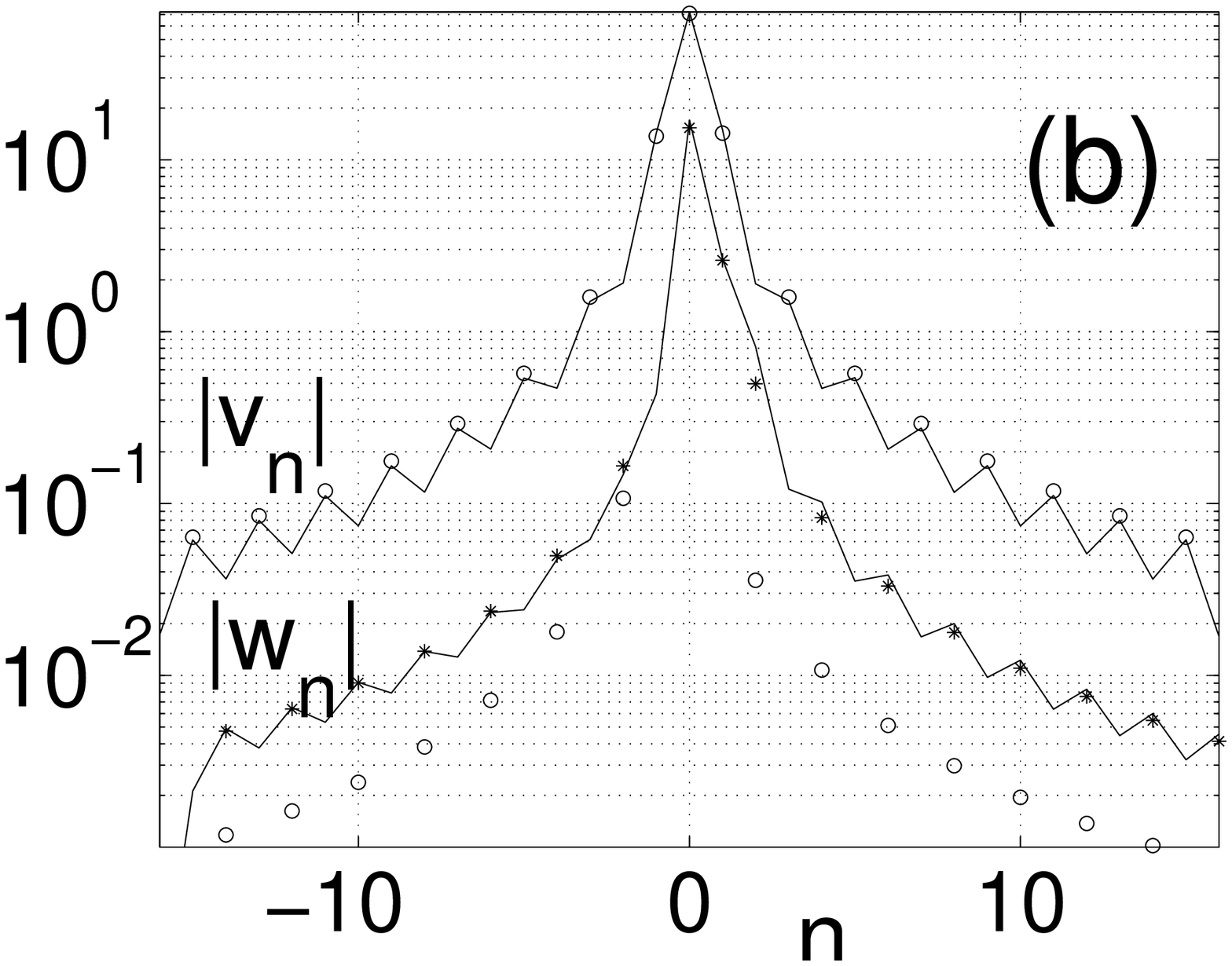,width=\figwidtha }}}}
\caption{(a) Ratio $R = |v_0|^2/|w_0|^2$ versus mismatch parameter $\sigma$ calculated exactly (asterisks) and by the averaged-field theory (solid).  (b) Fourier components of solution for $\kappa = 700$, $\sigma = 1$ calculated exactly (solid) and by the averaged-field theory (asterisks for $w_n$ and circles for $v_n$). $r = -1$.}
\label{pl}
\end{figure}

Figure \ref{pl} shows some of the exact solutions for $r = -1$ when there is both a linear and nonlinear grating present, with parameters: $a'/\kappa = 0.158$, $d_0'/d' = 5/3$ and $\eta = 1$.  These gratings correspond to the GaAs/GaAlAs structure reported in Ref.\ \onlinecite{Pet96}.   Figure \ref{pl}(a) shows the ratio $R = |v_0|^2/|w_0|^2$ versus the mismatch parameter $\sigma$ for grating wave numbers $\kappa = 700$ and $\kappa = 100$.  These values of $\kappa$ give $\tilde\gamma < 0.1$, which means that these solutions lie outside the stability region predicted by the average theory (Fig.\ \ref{stability}).  Not shown are the solutions at $\sigma = 0$ for which $\bar w$ is zero or negligibly small.  As the figures show, in terms of the DC intensity ratio, the exact solutions consistently agree well with the approximate solutions found by averaging.

Figure \ref{pl}(b) gives the Fourier coefficients $|w_n|$ and $|v_n|$ for a representative solution.  Even though the amplitudes do not decrease exponentially with $n$, the decrease is sufficiently rapid such that the tenth component is almost three orders of magnitude less than the DC value.  Thus in our calculations, we typically include up to the components $n=\pm 16$ components in the Fourier expansions.  

That the approximate theory accurately predicts the averaged properties of the system is well known.  However Fig.\ \ref{pl}(b) shows that this approach does not properly account for the higher-order Fourier components (particularly the odd components of $w$ and even components of $v$), which are important for a correct MI analysis             %

\section{EXACT STABILITY ANALYSIS}
\label{floquet_mi}
\subsection{Floquet theory}
We determine the linear stability of the plane-wave solutions $(w_s,v_s)$ by 
using exact Floquet theory
\cite{Fueetal97}.  For perturbations of transverse frequency offset $\nu$:
\begin{eqnarray}
  w(x,z) &=& w_s(z) + \delta_1(z)\exp{(-i\nu x)} + \delta_2^*(z)
  \exp{(i\nu x)}, \nonumber \\
  v(x,z) &=& v_s(z) + \delta_3(z)\exp{(-i\nu x)} + \delta_4^*(z)
  \exp{(i\nu x)}, 
\end{eqnarray} 
the linear evolution is governed by the linearized equation $\partial_z{\bf P} = M(z){\bf P}$,
where
% the vector ${\bf P}(z)$ and the matrix $M(z)$ are
\begin{eqnarray}
  {\bf P} = \left( \begin{array}{c}
  \delta_1\\ \delta_2\\ \delta_3\\ \delta_4 
  \end{array} \right),\;  M = i\left(\begin{array}{cccc}
  a & b & c^* & 0 \\ -b^* & -a & 0 & -c \\ 
  2c & 0 & d & 0 \\ 0 & -2c^* & 0 & -d \end{array}\right),
\end{eqnarray}
with $a = -\nu^2/2 - r$, $b = v_s(z)\chi(z)\exp{(i\kappa z)}$, $c = w_s(z)\chi(z)\exp{(-i\kappa z)}$, and $d = -\nu^2/4 - \sigma + \alpha'(z)$. 
Because of the periodicity in the coefficients $\alpha'$ and $\chi'$ and in the solutions $w_s$ and $v_s$, the matrix $M$ governing the growth of linear perturbations is also periodic, with period $z_p = 2\pi/|\kappa|$.  To determine the stability, we need only find the eigenvalues 
$\lambda_i$ of the mapping $\delta T$ of the solution over one period\cite{Fueetal97}: ${\bf P}(z+z_p)=\delta T{\bf P}(z)$.
 
Now for $\{{\bf P}(z)\}_i=\delta_{i,k}$ for some $k\in\{1,2,3,4\}$, the solution one period later ${\bf P}(z+z_p)$ will be the $k$th column of $\delta T$.  Thus we may construct the constant matrix $\delta T$ by integrating the linearized equation $\partial_z{\bf P} = M(z){\bf P}$ out to $z=z_0+z_p$ for the four initial conditions $ \{{\bf P}(z_0)\}_i =\delta_{i,k}$ corresponding to each $k \in \{1,2,3,4\}$.  This is done numerically with a fourth-order Runge-Kutta method.  If all the $\lambda_i$ lie on the unit circle for every 
$\nu$, then the solutions $w_s$ and $v_s$ are stable; otherwise  
there is a net gain whose profile is well-estimated by the maximum eigenvalue growth rate $ g=\max\{\Re[\ln(\lambda_i)/z_p]\}$.

\subsection{Applications I: Efficient phase matching}
\label{appI}
To help in the analysis of the Floquet results, we divide the situations into two regimes: efficient phase matching, where the residual effective mismatch is much smaller than the material mismatch ($\tilde \beta' \ll \beta'$), and poor phase matching.  In the former case, the averaged-field theory is expected to hold, whereas in the latter the large residual mismatch invalidates the assumptions made in the perturbation theory (i.e. $\tilde \beta ' \ll \kappa$).

We consider first the case which corresponds to having no effective quadratic nonlinearity in the averaged-field analysis (i.e. $\rho = 0$), to see whether the Floquet analysis confirms the strong prediction of the averaged-field theory that the solutions and their MI properties are governed by cubic nonlinearities.

\begin{figure}
  \centerline{\hbox{
  \psfig{figure=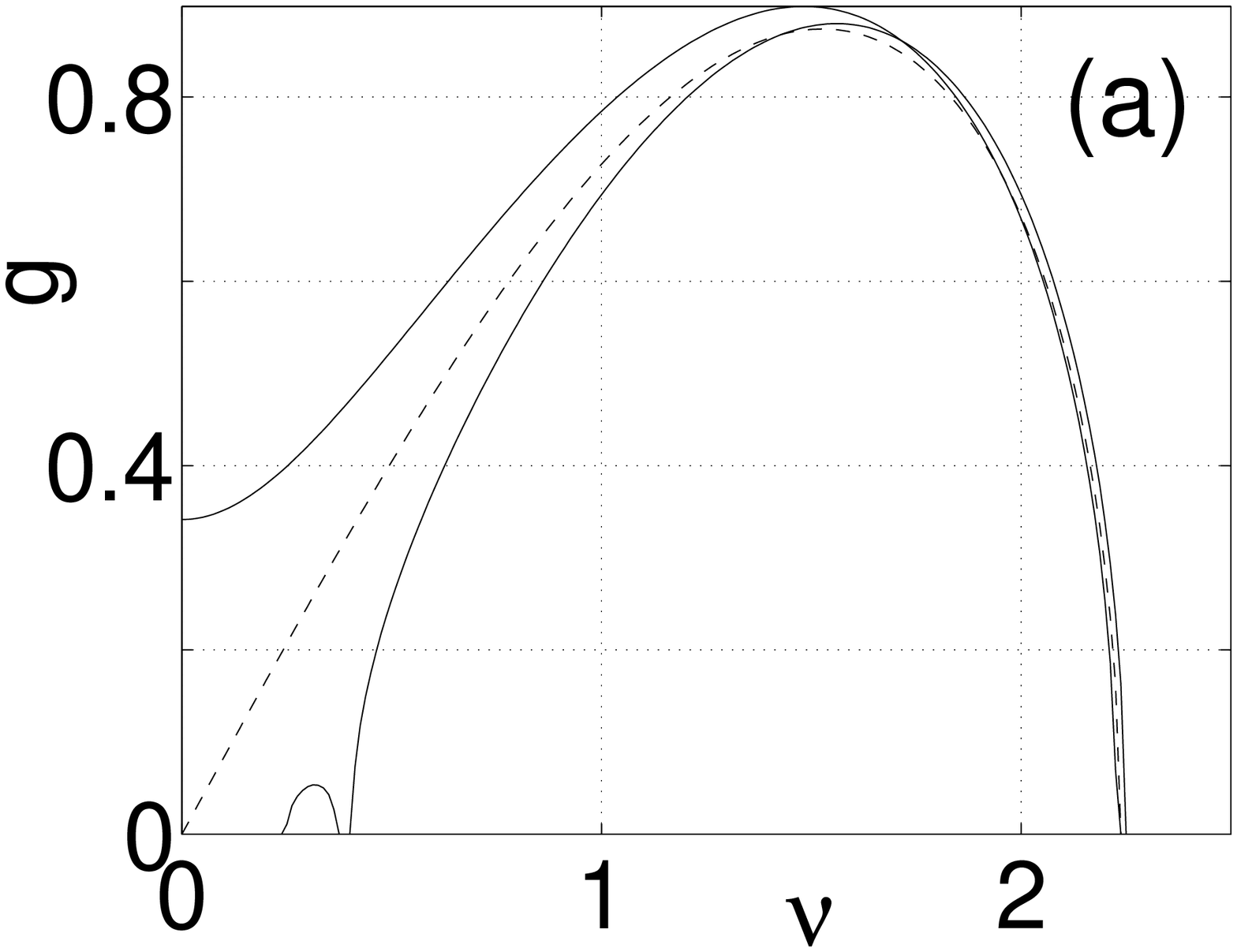,width=\figwidtha}
  \psfig{figure=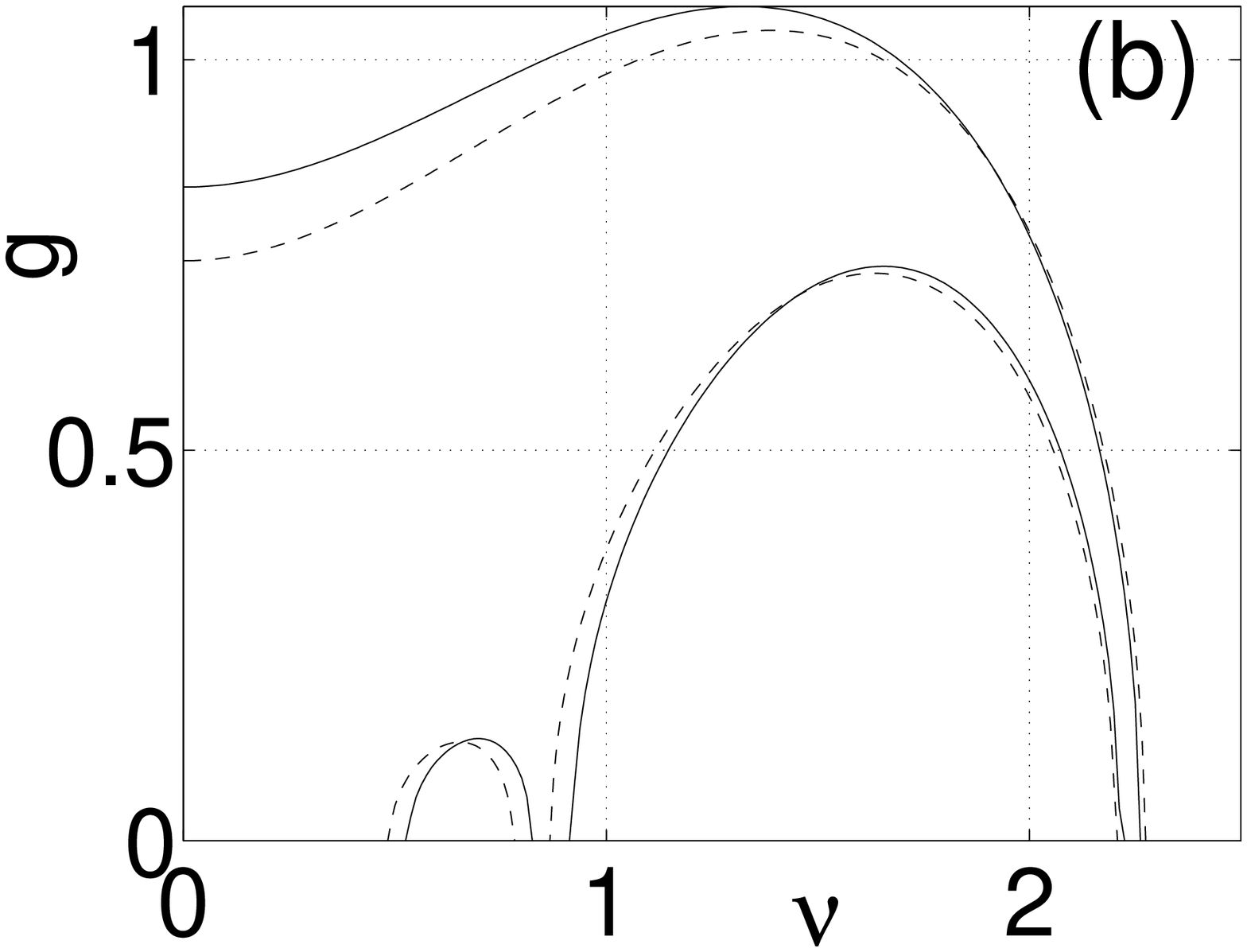,width=\figwidtha}}}
  \caption{Gain curves for the dual plane-wave solution when (a) $\rho = 0$ and (b) $|\rho| = 2/(15\pi)$, for $\gamma < 0$ and $\sigma = -0.5$ ($\tilde\beta' = -3/2$).  The dashed curve gives the averaged-field results, the solid curves give the exact Floquet results.}
  \label{gain_2}
\end{figure}

To make the comparison with the exact Floquet analysis, consider a physical system with the parameters: $a'/\kappa = 5/14$ and $d_0'/d' = 14/5$, for which $\rho = 0$.  
Figure \ref{gain_2}(a) shows the MI gain profiles (gain versus the spatial frequency $\nu$ of the perturbation) for the $\bar v_s \neq 0$ solution, corresponding to the unstable background in Fig.\ \ref{dark1}(b), calculated by both the exact Floquet theory (solid line) and the averaged-field theory (dashed). The Floquet theory predicts that each of the two solutions has its own distinct gain curve, whereas in the approximate calculation, the two solutions ($\pm \bar v$) give the same degenerate gain curve.  The approximate and exact results agree very well away from $\nu = 0$, but noticeably differ near $\nu = 0$.  For comparison, Figure \ref{gain_2}(b) plots the gain curve for $|\rho| = 2/(15\pi)$, using $a'/\kappa = 1/3$.  Here the Floquet curves for the two solutions differ more markedly, but each agrees well with its approximate counterpart, which now correspond to points on the (U2) and (G) branches in Fig.\ \ref{plane_solns}.   The discrepancy in Fig.\ \ref{gain_2}(a)in is due to higher-order terms neglected in the averaged theory which break the degeneracy of the NLS solution (\ref{2dark}).  These terms, of order $\epsilon^2$ and higher, would need to be included for consistent theory that accurately predicts the gain when $\nu$ is smaller than order $\epsilon^1$, and especially when the quadratic (order $\epsilon^0$) nonlinearity is absent. 

The $\bar v_s = 0$ solution is predicted by the averaged-field theory to be stable.  However the exact Floquet theory shows that there {\em is} some nonzero gain at high spatial frequencies.   However, the gain curves are extremely narrow and have low peak values: the largest region of gain at $\nu = 9.9$ is only $\Delta \nu = 0.05$ wide with a maximum of $g = 0.21$, which corresponds with the weak instability of the simulation in Fig.\ \ref{dark1}(a).  

\begin{figure}
  \centerline{\hbox{
  \psfig{figure=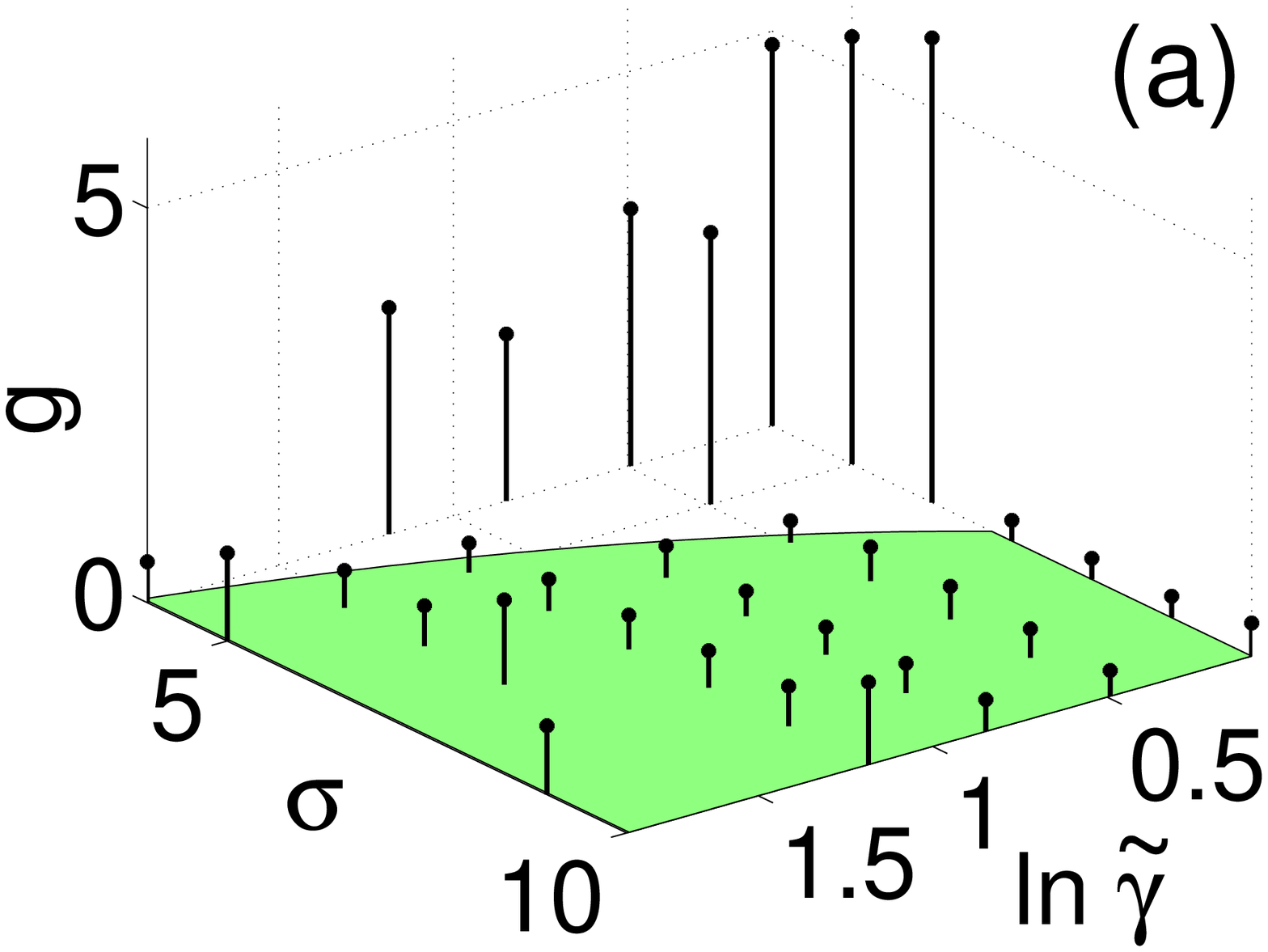,width=\figwidtha}
  \psfig{figure=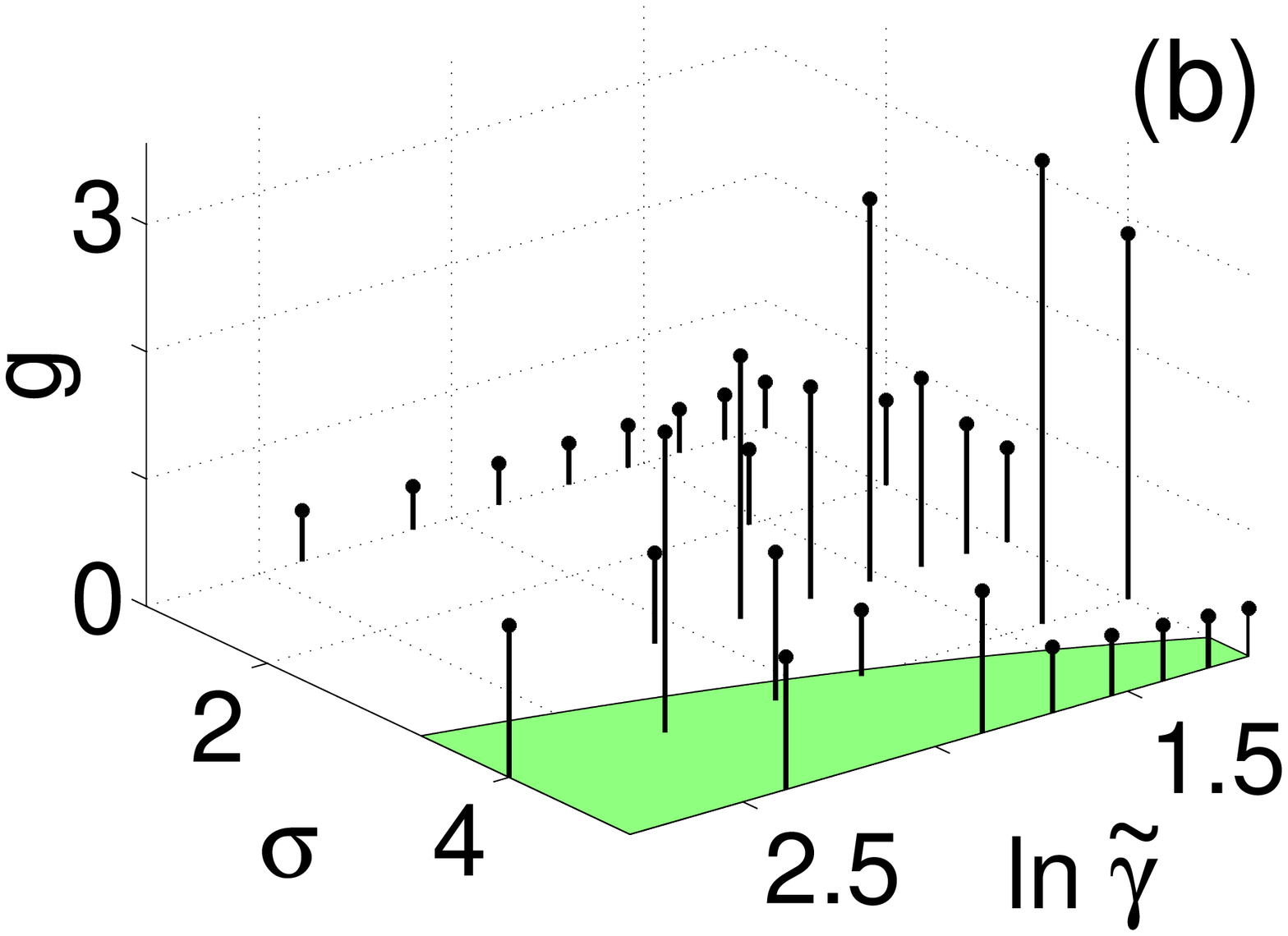,width=\figwidtha}}}
  \caption{Maximum gain for solutions with (a) $5 \le \kappa \le 30$ (b) $1 < \kappa< 10$.  The region of zero gain predicted from the averaged-field theory is shaded.    $d_0'/d' = 5/3$ and $a'/\kappa = 0.316$.}
  \label{pl_stab}
\end{figure}

A Floquet analysis of the solutions shown in Fig.\ \ref{pl} does not find any points that are stable.  This is not surprising, because these points lie outside the stability region predicted by the first-order analysis. But exact solutions corresponding to the stable lowest branch of averaged solutions can be found for lower values of $|\kappa|$, the grating wave number.  A plot of the maximum MI gain (\ref{pl_stab}(a)), however, shows that these solutions exhibit a weak, but nonzero MI gain.  This residual gain ($\simeq 0.2 \rightarrow 1$) increases for decreasing $\kappa$ and is caused by narrow gain bands that are once again present at high $\nu$.  Other simulations also confirm that such high-$\nu$ peaks, which are not predicted by the averaged theory, are a general feature of the gain spectrum for efficient QPM gratings.

\begin{figure}
  \centerline{\hbox{
  \psfig{figure=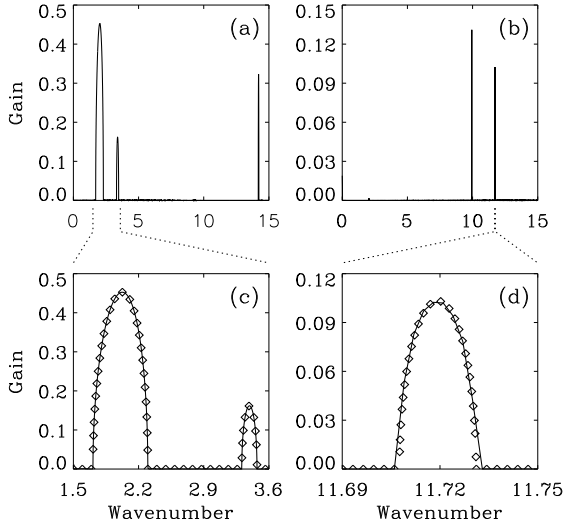,width=\figwidthc}}}
\vspace*{0.5cm}  
\caption{Floquet gain spectra for (a) the LiNbO$_3$ and (b) the GaAs/AlAs gratings, for $\sigma$=$-2$. The diamonds give the averaged-field results in (c) and the equivalent non-phase-matched homogeneous results in (d).}
  \label{gain_4}
\end{figure}

To analyze the MI spectrum in detail and illustrate the physical origin of these high-$\nu$ peaks, we consider the spectra of two concrete examples that are typical for efficiently phase-matched gratings ($|\tilde \beta| \ll |\beta|$).   Figure \ref{gain_4}(a) is the spectrum for a conventional domain-reversal grating in LiNbO$_3$ ($a'=d_0'=0$) for $r$=$-1$, $\kappa$= 100 and exact phase matching ($\sigma$=$-2$), and Fig.\ \ref{gain_4}(b) is that of a GaAs/AlAs superstructure with a nonlinear grating etched though quantum-well disordering\cite{HelHutetal00} ($d_0'/d'$=4.6) and a linear grating chosen to be $a'/\kappa$=$10/46$.  The former is unstable according to the averaged-field analysis and possesses a low-$\nu$ gain that is accurately predicted by the average theory (Fig.\ \ref{gain_4}(c)); the latter lies in the predicted stable region and thus has no gain at low $\nu$.

The high-$\nu$ gain bands are related to the inherent MI in the non-phase-matched, homogeneous $\chi^{(2)}$ material (i.e. with no grating), as demonstrated by the close match between the homogeneous gain band (Fig.\ \ref{gain_4}(d)) and one of the peaks in Fig.\ \ref{gain_4}(b).  In general, each gain peak that appears in the homogeneous spectrum usually also appears in the Floquet spectrum, typically with the same height and spectral location.  Often, as in Fig.\ \ref{gain_4}(b), the `homogeneous' peak is split into several closely spaced secondary peaks by the grating.  Sometimes, as for a symmetric grating ($d_0'$=0) such as in LiNbO$_3$ (Fig.\ \ref{gain_4}(a)), the original homogeneous peak is suppressed, leaving only the secondary peaks in the high-$\nu$ part of the spectrum.

\subsection{Applications II: Poor phase matching}
As we saw in Sec.\ \ref{average_mi}, attempting to reach the stable branch of the approximate solutions while maintaining a large quadratic nonlinearity ($\rho$) usually involves the poor-phase-matching regime, where the averaged theory no longer holds.  To illustrate what happens, we show in Fig.\ \ref{pl_stab}(b) the maximum growth rate of instabilities for the grating  $d_0'/d' = 5/3$ and $a'/\kappa = 0.316$. Here $-\tilde\beta'> 3$ is rather large and $\kappa< 10$ is fairly small, so the first-order perturbation results are not guaranteed to be valid in this region.  That the Floquet theory found solutions with large gain in the predicted stable region is therefore not surprising. 

\begin{figure}
  \centerline{\hbox{
  \psfig{figure=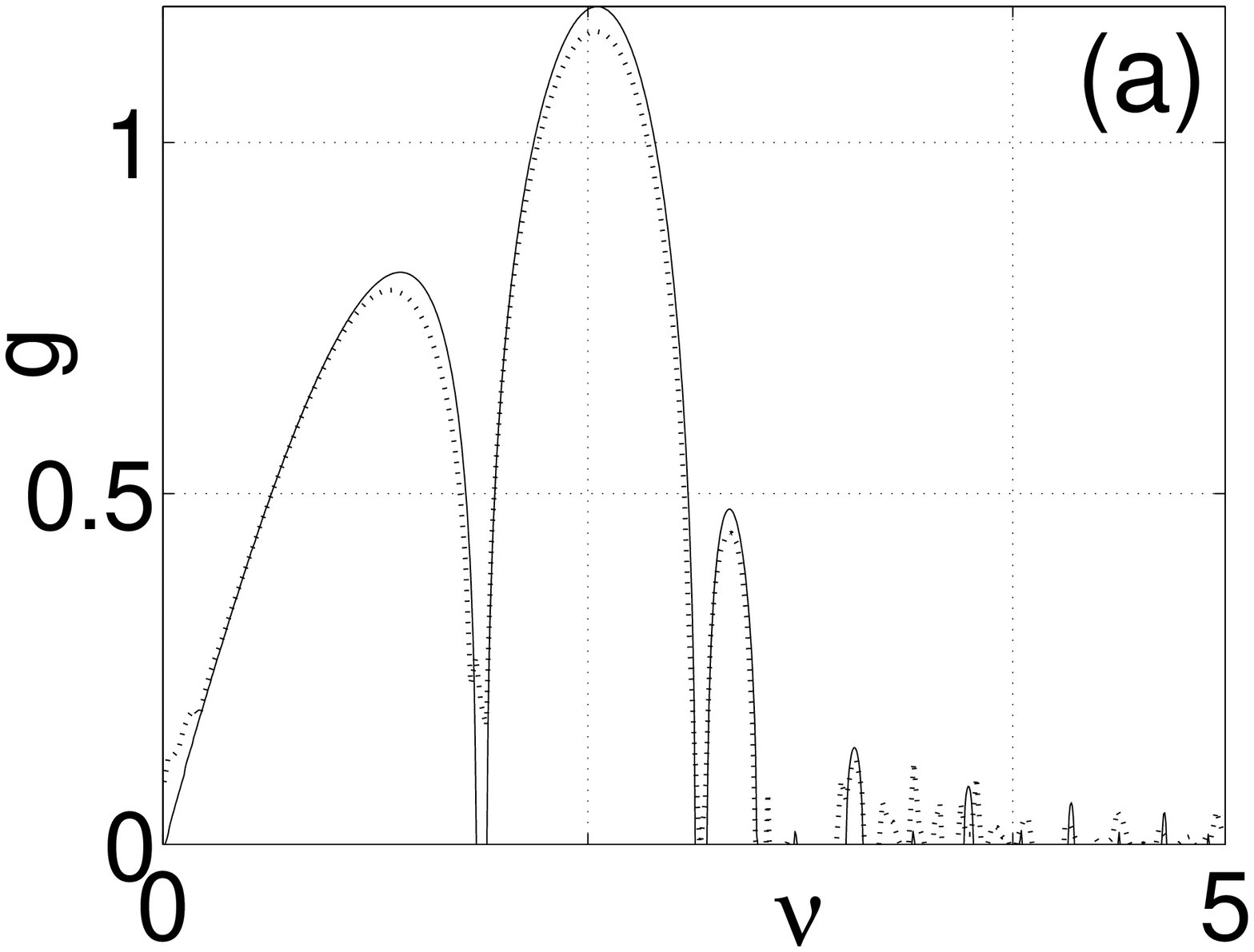,width=\figwidtha}
  \psfig{figure=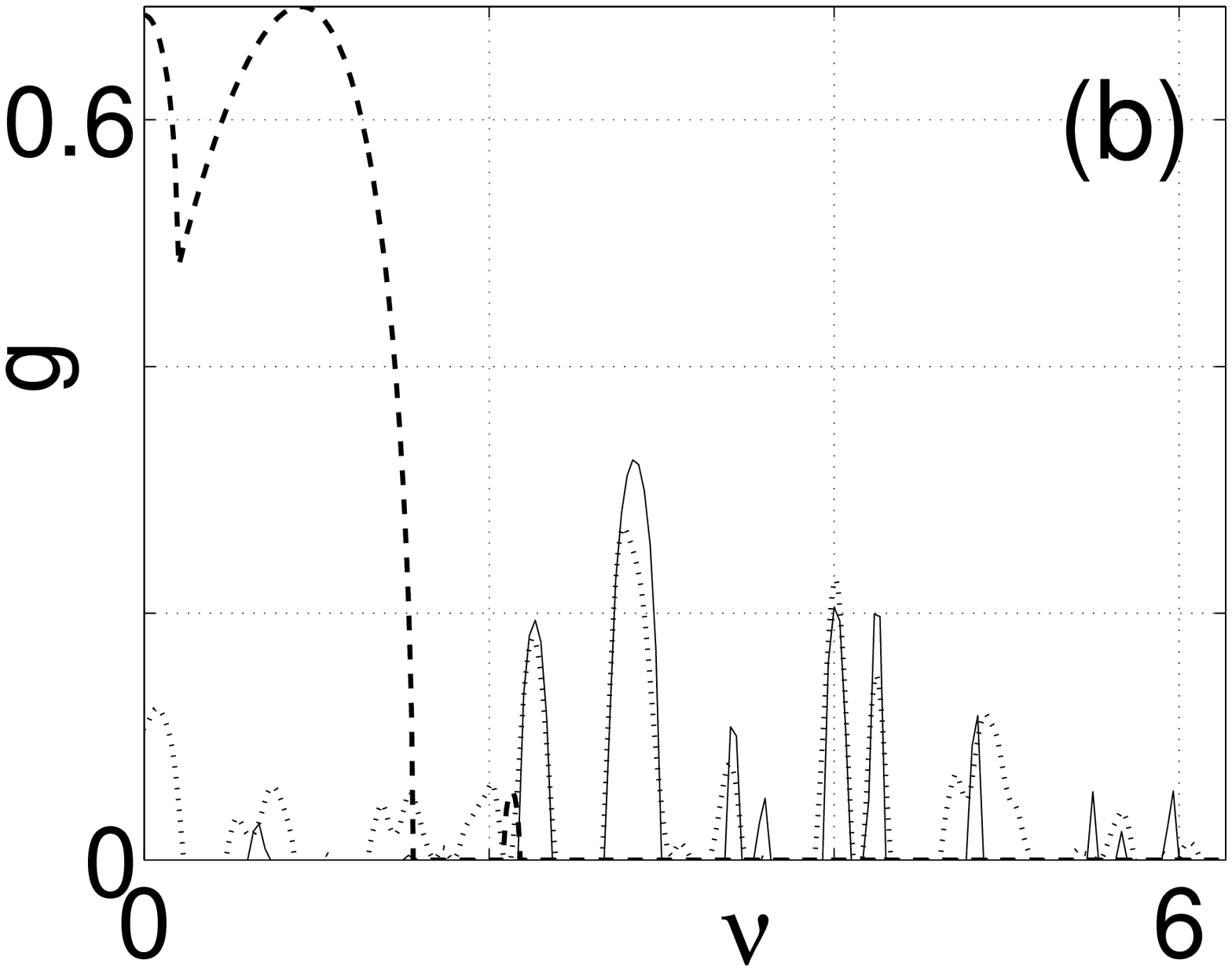,width=\figwidtha}}}
  \caption{Gain curve for (a) $\kappa = 2$ and $\sigma = 4$ and (b) $\kappa = 6$ and $\sigma = 1$, calculated from  Floquet theory (solid) and simulations (dotted).  The averaged theory predicts a stable solution in (a) and a gain marked by the dashed line in (b).  $d_0'/d' = 5/3$ and $a'/\kappa = 0.316$.}
  \label{gain_curve2}
\end{figure}

The differences between the averaged and exact treatments for these parameters are further illustrated by comparison of the gain profiles, which we do in Fig.\ \ref{gain_curve2} for two representative points.  The averaged solution corresponding to Fig.\ \ref{gain_curve2}(a) ($\kappa = 2$, $\sigma = 4$) lies in the predicted stable region, but the Floquet analysis gives a nonzero gain which is large and broadband.  The exact gain curve in Figs.\ \ref{gain_curve2}(b) ($\kappa = 6$, $\sigma = 1$) consists of a number of narrow peaks distributed over a range of frequencies, whereas the first-order analysis predicts a larger, broadband curve that lies at low frequencies.
 
In Sec.\ \ref{appI}, where the grating efficiently phase-matches the fields, we saw that the MI gain spectrum falls into two, well-separated parts: the low-frequency bands induced by the periodicity and accurately predicted by the averaged equations, and the high-frequency bands related to inherent $\chi^{(2)}$ gain.  When the residual mismatch $|\tilde \beta|$ is larger (and not so different from $|\beta|$), we find that the two regions no longer are well separated and that the averaged equations no longer accurately predict the low-$\nu$ bands.  For very poor phase matching as in this section, the two features mix, producing a complicated gain spectrum (e.g. Fig.\ \ref{gain_curve2}(b)) that cannot be simply related to the distinct underlying physical mechanisms.

\subsection{Experimental Stability}
\label{exp}
In all our results, we found no solutions that, mathematically speaking, were modulationally stable.  Even in the region where the averaged-field analysis predicts stability, there is a residual gain corresponding to the high-$\nu$ bands.  However this residual gain is not large ($g \le 0.3$ or smaller for material mismatch $|\beta'| \sim 10$ or larger), and may be neglected under a reasonable definition of {\em experimental stability}.  If perturbations do not increase by any orders of magnitude within a few diffractions lengths, which is a typical experimental length scale, then the instability will not be detectable.  However if we do neglect these high-$\nu$ bands, so that the averaged and Floquet results coincide, then we must also allow some solutions previously classed as unstable by the averaged-field analysis to be classed as stable, because the low-$\nu$ bands are sometimes smaller than the high-$\nu$ bands. These additional solutions correspond to the light-grey regions in Fig.\ \ref{unstable} for $r = -1$ as well as similar regions for $r = 0$ and $r = 1$, which were previously regarded as totally unstable branches.

\section{PROPAGATIVE SIMULATIONS}
\label{propagative}
To confirm the Floquet predictions of the MI spectra, we perform numerical simulations of the (scaled) field equations (\ref{planewave2}).  We can determine gain curves by launching plane waves seeded with small fluctuations, which then evolve linearly.  Using Gaussian white noise to excite all frequencies, we calculate the gain curve from the Fourier transformed amplitude\cite{HeArretal99} $\tilde E_1$ as: $ \tilde g(\nu) = [\ln|\tilde E_1(Z_2,\Upsilon)| - \ln|\tilde E_1(Z_1,\Upsilon)|]/(Z_2-Z_1)$, where $\Upsilon = \sqrt{\eta} \nu$.

\begin{figure}
  \centerline{\hbox{
  \psfig{figure=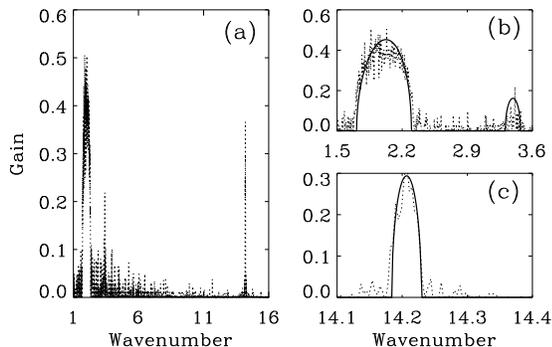,width=\figwidthc}}}
  \vspace*{0.6cm}
  \caption{(a) Gain profile calculated from propagative simulations, for the same parameters as in Fig.\ \ref{gain_4}(a). Comparisons with the Floquet theory (solid curve) are given in (b) and (c). }
 \label{prop3}
\end{figure}

Propagation calculations confirm, for example, the gain curves predicted by Floquet theory in Fig.\ \ref{gain_2}(a).  Importantly, they agree with the Floquet curves in regions where the averaged-field and Floquet theories predict different results, such as near $\nu = 0$ in these figures.   A more dramatic illustration of this is shown in Figs.\ \ref{gain_curve2}, where the averaged and exact predictions markedly disagree.  Small, narrow peaks are difficult to distinguish from the noise floor in these simulations, but the peaks seen in the smoothed data in Fig.\ \ref{gain_curve2}(b) (dotted) clearly correspond to the largest Floquet peaks (solid).  For the same reason, the small, narrow high-$\nu$ bands predicted by Floquet theory are difficult to detect, but with a careful choice of simulation parameters these peaks can be seen, as shown in Fig.\ \ref{prop3} for the LiNbO$_3$ grating.  Clearly, the simulations confirm the Floquet calculations, in both the low-$\nu$ [Fig.\ \ref{prop3}(b)] and high-$\nu$ [Fig.\ \ref{prop3}(c)] regions.

\section{CONCLUSIONS}
In summary, we determined the exact structure and modulational stability of plane-wave solutions in periodic $\chi^{(2)}$ materials, using both Floquet transfer-matrix calculations and direct numerical simulations.  In particular, we found Fourier expansions of the plane waves in the propagation direction and used these in calculations of the gain spectrum (growth rate versus transverse frequency of the instability).  In the regime of efficient phase matching, we found that, due to the periodicity, the spectra contain gain bands both at low spatial frequency, which are caused by the phase-matching gratings, and at high spatial frequency, which are related to the inherent gain of homogeneous $\chi^{(2)}$ materials.  For poor phase-matching, the gain bands do not fall into these different regions of the spectrum and so cannot be respectively attributed to distinct underlying causes. 

In the regime of efficient phase matching, the exact calculations were supplemented by a simpler averaged-field approach characterized by effective quadratic and induced cubic nonlinearities.  For sufficiently large cubic nonlinearities and away from exact phase matching, the averaged-field theory predicts stability for some solutions.  However, the exact techniques showed that whereas the averaged-field approach accurately predicts slowly varying properties of the system, it sometimes inaccurately treats the higher Fourier components, as these are treated to a lower order in the theory. It also shows inaccuracies in the MI predictions, being unable to predict the high-frequency gain bands.  Nevertheless, the averaged-field theory accurately describes low-frequency gain, even in the extreme case where the effective quadratic nonlinearity is cancelled by competition between the linear and nonlinear gratings.  Futhermore, because the high-frequency gain is often small, the averaged-theory zero-gain predictions can be accurate under experimentally relevant definitions of stability.

\vspace*{-5 mm}\acknowledgments{This work was supported by the Danish Technical Research Council under grant 26-00-0355.}

\end{multicols}  %Remove before submission

\vspace*{-5 mm}\begin{thebibliography}{}

\vspace*{-1.5 cm}\bibitem{SteHagTor96}
G. I. Stegeman, D. J.  Hagan and L.  Torner,
  ``$\chi^{(2)}$ cascading phenomena and their applications to all-optical signal processing, mode-locking, pulse compression and solitons,''
Opt.  Quantum Electron.  {\bf 28}, 1691--1740 (1996). 

\bibitem{HayKos93}
K. Hayata and M. Koshiba,
``Multidimensional solitons in quadratic nonlinear media,''
\prl {\bf 71}, 3275--3278 (1993).

\bibitem{TorMenTorSte95}
L. Torner, C. R. Menyuk, W. E. Torruellas and G. I. Stegeman,
``Two-dimensional solitons with second-order nonlinearities,''
\ol {\bf 20}, 13--15 (1995).

\bibitem{BerMezRasWyl95}
L. Berg\'e, V. K.  Mezentsev, J. J. Rasmussen and J. Wyller,
 ``Formation of stable solitons in quadratic nonlinear media'',
\pra {\bf 52}, 28--31 (1995).

\bibitem{PelBurKiv95}
D. E. Pelinovsky, A. V. Buryak and Y. S. Kivshar,
``Instability of solitons governed by quadratic nonlinearities'',
\prl {\bf 75}, 591--595 (1995).

\bibitem{MalDruetal97}
B. A. Malomed, P. D. Drummond, H. He, A. Berntson, D. Anderson and M. Lisak,
``Spatiotemporal solitons in multidimensional optical media with a quadratic nonlinearity,''
Phys.\ Rev.\ E {\bf 56}, 4725--4735 (1997).

\bibitem{SchBaeSte96}
R. Schiek, Y. Baek and G. I. Stegeman,
``One-dimensional spatial solitary waves due to cascaded second-order nonlinearities in planar waveguides,''
Phys.\ Rev.\ E {\bf 53}, 1138--1141 (1996).

\bibitem{TorWanetal95}
W. E. Torruellas, Z. Wang, D. J. Hagan, E. W. Van Stryland, G. I. Stegeman, L. Torner and C. R. Menyuk,
``Observation of two-dimensional spatial solitary waves in a quadratic medium,''
\prl {\bf 74}, 5036--5039 (1995).

\bibitem{LiuBecWis00}
X. Liu, K. Beckwitt and F. Wise,
``Two-dimensional optical spatiotemporal solitons in quadratic media,''
Phys.\ Rev.\ E {\bf 62}, 1328--1340 (2000)

\bibitem{LopCouetal01}
E. L\'opez-Lago, V. Couderc, C. de Angelis, F. Gringoli and A. Barth\'el\'emy,
``Experimental demonstration of the self-trapping of a weak probe induced by a quadratic spatial soliton,''
in {\em Nonlinear Guided Waves and Their Applications}, OSA Technical Digest (Optical Society of America, Washington DC, 2001), pp 379--381.

\bibitem{CouLopSimBar01}
V. Couderc, E. L\'opez-Lago, C. Simos and A. Barth\'el\'emy,
``Experiments in quadratic spatial soliton generation and steering in a noncollinear geometry,''
\ol {\bf 26}, 905--907 (2001).

\bibitem{TorAssetal96}
W. E. Torruellas, G. Assanto, B. L. Lawrence, R. A. Fuerst and G. I. Stegeman,
``All-optical switching by spatial walkoff compensation and solitary-wave locking,''
App. Phys. Lett. {\bf 68}, 1449-1451.

\bibitem{DiTChietal00}
P. Di Trapani, W. Chinaglia, S. Minardi, A. Piskarskas and G. Valiulis,  ``Observation of quadratic Optical Vortex Solitons,'' \prl {\bf 84}, 3843--3846 (2000).
 
\bibitem{TriFer95}
S. Trillo and P.  Ferro, ``Modulational instability in second-harmonic generation,''  \ol {\bf 20}, 438--440 (1995). 

\bibitem{Fueetal97}
R.~A. Fuerst, D.-M. Baboiu, B. Lawrence, W.~E. Torruellas, G.~I. Stegeman, S. Trillo and S. Wabnitz, ``Spatial modulational instability and multisolitonlike generation in a quadratically nonlinear optical medium'', Phys. Rev. Lett. {\bf 78}, 2756--2759 (1997).                            

\bibitem{FanMalSchSte00}
H. Fang, R. Malendevich, R. Schiek and G. I. Stegeman,
``Spatial modulational instability in one-dimensional lithium niobate slab waveguides,''
\ol {\bf 25}, 1786--1788 (2000).

\bibitem{SchFanMalSte01}
R. Schiek, H. Fang, R. Malendevich and G. I. Stegeman,
``Measurement of modulational instability gain of second-order nonlinear optical eigenmodes in a one-dimensional system.'',
\prl {\bf 86}, 4528--4531 (2001).

\bibitem{BanKivBur97}
O. Bang, Y. S. Kivshar and A. Buryak, ``Bright spatial solitons in defocusing Kerr media supported by cascaded nonlinearities,'' \ol {\bf 22}, 1680-1682 (1997).

\bibitem{AleBurKiv98}
T.~J. Alexander, A.~V. Buryak and Y.~S. Kivshar, ``Stabilization of dark and vortex parametric spatial solitons,'' \ol {\bf 23}, 670--672 (1998).

\bibitem{ClaBanKiv97}
C.~B. Clausen, O. Bang and Y.~S. Kivshar, ``Spatial solitons and induced Kerr effects in quasi-phase-matched quadratic media,'' \prl {\bf 78}, 4749--4752 (1997). 

\bibitem{CorBan01a}
J.~F. Corney and O. Bang, ``Modulational stability in periodic quadratic nonlinear materials,''
\prl (to appear).

\bibitem{BanBalChrTor99}
O. Bang, C.~B. Clausen, P.~L. Christiansen and L. Torner, ``Engineering competing nonlinearities,'' \ol {\bf 24}, 1413--1415 (1999).

\bibitem{CorBan01}
J.~F. Corney and O. Bang, ``Solitons in quadratic nonlinear photonic crystals,'' Phys.\ Rev.\ E (to appear).

\bibitem{DiTBraetal01}
P. Di Trapani, A. Bramati, S. Minardi, W. Chinaglia, S. Trillo, C. Conti, J. Kilius and G. Valiulis,
``Focusing versus defocusing nonlinearities in self-trapping due to parametric frequency conversion,''
\prl (to appear).

\bibitem{FejMagJunBye92}
M.~M. Fejer, G.~A. Magel, D.~H. Jundt, and R.~L. Byer, ``Quasi-phase-matched second harmonic generation: Tuning and tolerances,'' {IEEE} Journal of Quantum Electronics {\bf 28}, 2631--2654 (1992).

\bibitem{SuhNis90}
T. Suhara and H. Nishihara, ``Theoretical analysis of waveguide second-harmonic
 generation phase matched with uniform and chirped gratings,'' {IEEE} Journal of Quantum Electronics {\bf 26}, 1265--1276 (1990).
 
\bibitem{JasArvLau86}
B. Jaskorzynska, G. Arvidsson, and F. Laurell, ``Periodic structures for phase-matching in second harmonic generation in titanium lithium niobate waveguides,'' in {\it Spie: Integrated Optical Circuit Engineering {III}} {\bf 251}, 221--228 (1986).

\bibitem{TanBey73}
C. L. Tang and P. P. Bey,  ``Phase matching in second-harmonic generation using artificial periodic structure,'' {IEEE} J. Quantum Electron. {\bf 9}, 9--17 (1973).

\bibitem{Pet96}
D.~V. Petrov, ``Nonlinear phase shift by cascaded quasi-phase-matched second harmonic generation,'' \oc {\bf 131}, 102--106 (1996).

\bibitem{KobFedBanKiv98}
A. Kobyakov, F. Lederer, O. Bang and Y. S. Kivshar,
``Nonlinear phase shift and all-optical switching in quasi-phase-matched quadratic media,'' \ol {\bf 23}, 506--508 (1998).

\bibitem{BanGraCor01}
O. Bang, T. W. Graversen and J. F. Corney, ``Accurate switching intensities and length scales in quasi-phase-matched materials,'' \ol {\bf 26}, 1007-1009 (2001).

\bibitem{ArmBloDucPer62}
J.~A. Armstrong, N. Bloembergen, J. Ducuing and P.~S. Pershan, ``Interactions between light waves in a nonlinear dielectric,'' Phys. Rev. {\bf 127}, 1918--1939 (1962).

\bibitem{MenSchTor94}
C. R. Menyuk, R. Schiek and L. Torner, ``Solitary waves due to $\chi^{(2)}:\chi^{(2)}$ cascading,'' \josab {\bf 11}, 2434--2443 (1994).
   
\bibitem{Ban97} O. Bang, ``Dynamical equations for wave packets in material with both quadratic and cubic response,'' \josab {\bf 14}, 51--61 (1997).
 
\bibitem{HeDruMal96} 
H. He, P.~D. Drummond and B.~A. Malomed, ``Modulational stability in dispersive optical systems with cascaded nonlinearity,'' \oc {\bf 123}, 394--402 (1996).

\bibitem{HeArretal99} 
H. He, A. Arraf, C.~M. de Sterke, P.~D. Drummond and B.~A. Malomed, ``Theory of Modulational instability in Bragg gratings with quadratic nonlinearity,'' Phys.\ Rev.\ E {\bf 59}, 6064--6078 (1999).
    
\bibitem{HelHutetal00}
A.  Saher Helmy {\em et al.},
%A. Saher Helmy, D. C. Hutchings, T. C. Kleckner, J. H. Marsh, A. C. Bryce, J. M. Arnold, C. R. Stanley, J. S. Aitchison, C. T. A. Brown, K. Moutzouris and M. Ebrahimzadeh,
``Quasi phase matching in GaAs-AlAs superlattice waveguides through bandgap tuning by use of quantum-well intermixing",
\ol {\bf 25}, 1370--1372 (2000). 

\bibitem{CreDauRufTor98}
T. Cretegny, T. Dauxois, S. Ruffo and A. Torcini, ``Localization and equipartition of energy in the $\beta$-FPU chain: Chaotic breathers'', Physica D {\bf 121}, 109--126 (1998).
\end{thebibliography}
\end{document}